\documentclass[12pt,aps,prd,superscriptaddress,showpacs,longbibliography,floatfix,nofootinbib]{revtex4-2}

\usepackage[utf8]{inputenc}
\usepackage{slashed}
\pdfoutput=1

\usepackage{color}
\usepackage{graphicx}   
\usepackage{bm}
\usepackage{amsmath}
\usepackage{amsfonts}
\usepackage{hyperref}

\def\be{\begin{equation}}
\def\ee{\end{equation}}
\def\ba{\begin{eqnarray}}
\def\ea{\end{eqnarray}}

\begin{document}

\title{Quantum Field Theory in Large N Wonderland: Three Lectures}

\author{Paul Romatschke}
\affiliation{Department of Physics, University of Colorado, Boulder, Colorado 80309, USA}
\affiliation{Center for Theory of Quantum Matter, University of Colorado, Boulder, Colorado 80309, USA}

\begin{abstract}
In these lecture notes, I review how to use large N techniques to solve quantum field theories in various dimensions. In particular, the case of N-dimensional quantum mechanics, non-relativistic cold and dense neutron matter, and scalar field theory in four dimensions are covered. A recurring theme is that large N solutions are fully non-perturbative, and can be used to reliably access quantum field theory for parameter regions where weak-coupling expansions simply fail. 
\end{abstract}

\maketitle

\section{Preface}

You may have heard that Quantum Field Theory is a well-developed, mature discipline.

That all the easy problems have been done long ago.

That there is nothing left to discover.

That is not true.

Welcome to QFT in large N wonderland!

\section{Introduction}

The aim of these lecture notes is to provide an accessible introduction to the physical applications of large N solution techniques for quantum field theory. They are aimed at advanced graduate students and early-career postdocs in theoretical physics, but they do contain new material that occasionally puzzles more senior researchers. The main guiding principle behind the lectures is that they offer techniques to obtain direct first-principle quantitative answers to physics problems of interest, with minimal specialized mathematical knowledge.

To keep the lecture notes readable, I have chosen to keep references at a minimum, with an emphasis on recent rather than older results.

That said, the use of large N techniques as opposed to perturbation theory has a long history in field theory, with many of the key results already obtained in the 1970s \cite{tHooft:1973alw,Abbott:1975bn,Linde:1976qh}. Unfortunately, subsequent research showed that large N techniques are not sufficient to solve specific non-abelian gauge theories of interest, such as QCD. As a consequence, the main theoretical tools for the study of QCD presently are perturbative (weak-coupling) expansions (e.g. \cite{vanRitbergen:1997va,Kajantie:2002wa,Gorda:2023mkk}), lattice QCD (e.g.\cite{Borsanyi:2020mff,BMW:2008jgk}), as well as effective non-relativistic expansions (e.g. \cite{Brambilla:2010cs}).

In contradistinction to QCD, large N does play an important role in holographic conjectures of supersymmetric gauge theories, such as the conjectured dual of ${\cal N}=4$ Super-Yang-Mills theory in the large N limit to classical Einstein gravity in asymptotically $AdS_5$ spacetimes \cite{Maldacena:1997re}. 

Coming full circle, the holographic conjectures for large N gauge theories did lead to conjectures for large N scalar theories, such as the conjectured dual of the O(N) model in 3 dimensions to higher-spin gravity in asymptotically $AdS_4$ spacetimes \cite{Klebanov:2002ja}. Unlike the case of gauge theories, where a proof of the gravity dual seems out of reach, the solvability of scalar field theories in the large N limit suggests the gravity dual theory can be derived, rather than conjectured \cite{Aoki:2016ohw,deMelloKoch:2018ivk,Aharony:2020omh}.

Despite the attractive feature of large N solvability, applications of large N techniques for scalar and fermionic theories has remained somewhat dormant since the 1970s. This provides opportunity for using large N techniques to solve problems of interest, such as calculation of transport coefficients \cite{Moore:2001fga,Aarts:2005vc,Romatschke:2021imm}, finite temperature correlators \cite{Romatschke:2019ybu,Weiner:2022kgx}, finite-density systems \cite{Lawrence:2022vwa}, as well as real-time evolution in quantum field theory \cite{Aarts:2006cv}.

Many problems which are intractable using standard perturbation theory surprisingly become not only possible but easy using large N expansions. This is a consequence of using $\frac{1}{N}$ as the  small expansion parameter, which allows direct access to observables for any value of the coupling, small or large.

The cutting edge application of this technique is four-dimensional theories, where a combination of large N expansion and non-perturbative renormalization techniques \cite{Parisi:1975im} allow one to circumvent long-held convictions about quantum triviality and asymptotically free theories. A surprising consequence is the possibility of a discovering a simpler version of Standard Model of Physics, with fewer parameters than needed for the Higgs mechanism.

Last but not least, working with large N expansions is fun! It directly combines relatively straightforward math with physics observables of interest, and new discoveries seem to be waiting at (almost) every page of a calculational notebook! Best of all, most of the large N wonderland is essentially unexplored, so that you can still claim your own patch (or country, or continent, or planet) in it!

Welcome to Quantum Field Theory in Large N Wonderland!

\section{Lecture 1: Quantum Mechanics}

Let's start with a simple test case where we can check our methods: quantum mechanics.

Quantum mechanics concerns itself with the spectrum of a Hamiltonian. For concreteness, let us consider the case of a one-dimensional system with Hamiltonian
\be
\label{ham1}
{\cal H}=\frac{p^2}{2}+\lambda x^4\,,
\ee
where $p,x$ are the momentum and position operator, respectively. The spectrum $E_n$ of the Hamiltonian is defined through the time-independent Schr\"odinger equation,
\be
\label{ev1}
\langle x | {\cal H} | n \rangle = {\cal H} \psi_n(x) = E_n \psi_n(x)\,,
\ee
where $\psi_n(x)$ are the wave-function eigenstates of ${\cal H}$.

What is the ground-state energy $E_0$ for the Hamiltonian (\ref{ham1}) ?

It so happens that $E_0$ for the Hamiltonian (\ref{ham1}) is not known analytically. I've chosen (\ref{ham1}) deliberately partly because of this property, otherwise it would be too easy. However, note that \textit{no} Hamiltonian with potential $V(x)\propto x^\alpha$ for $\alpha \in (2,\infty)$ has analytically known ground state energies, so the problem of finding $E_0$ is not contrived, but rather generic.

However, (\ref{ham1}) shares certain important properties with Hamiltonian of the harmonic oscillator $V(x)\propto x^2$, in that it's spectrum for $\lambda>0$ is real, discrete, and positive definite. It's just hard to calculate $E_0$.

Since our goal is to learn something about quantum field theory rather than quantum mechanics, let's cast quantum mechanics into field theory language by using path integrals. A rigorous way to do this from first principles is to consider the canonical partition function
\be
\label{z0}
Z={\rm Tr} e^{-\beta {\cal H}}=\sum_{n=0}^\infty \langle n | e ^{-\beta {\cal H}}|n\rangle = \sum_{n=0}^\infty e^{-\beta E_n}\,,
\ee
where $\beta=\frac{1}{T}$ and $T$ the temperature of the system. By inserting complete sets of states, one can turn the trace of the Boltzmann operator into a path integral (see the steps leading from (1.27) to (1.37) in the excellent open-access textbook \cite{Laine:2016hma})
\be
Z=\int {\cal D}\phi e^{-S_E}\,, \quad S_E=\int_0^\beta \left[\frac{1}{2}\dot\phi^2(\tau)+\lambda \phi^4(\tau)\right]\,,
\label{z1}
\ee
where $S_E$ is the Euclidean action of the theory and the field $\phi(\tau)$ lives on the Euclidean circle $\tau \in [0,\beta]$ with periodic boundary conditions.

Unfortunately, trying to solve the path integral in (\ref{z1}) is just as hard (or maybe even harder) than trying to directly solve the eigenvalue problem (\ref{ev1}). Some new idea is needed.

To develop this idea, let's do something counter-intuitive: instead of considering the quantum mechanics problem in one dimension (which was hard), how about quantum mechanics in higher dimensions? At first glance, if the problem was hard in one dimension, it seems unlikely that could make progress by trying to to solve it in two, three, etc. dimensions, but let's see.

Using the odd symbol $N$ to denote the number of dimensions, the equivalent Hamiltonian to (\ref{ham1}) is given by
\be
\label{ham2}
{\cal H}=\frac{\vec{p}^{\, 2}}{2}+\frac{\lambda}{N} \left(\vec{x}^2\right)^2\,,
\ee
where $\vec{p}=\left(p_1,p_2,\ldots p_N\right)$ and $\vec{x}=\left(x_1,x_2,\ldots,x_N\right)$ are again the momentum and position operators for quantum mechanics in $N$ dimensions. The appearance of $N$ in the denominator of the coupling $\lambda$ may appear arbitrary at first sight, but if one considers that $\vec{x}^{\,2}=x_1^2+x_2^2+\ldots x_N^2$ are $N$ contributions of the operator $x^2$ it becomes clear that $\frac{\lambda}{N}$ is the right normalization so that ${\cal H}$ scales appropriately with $N$. (Alternatively, or rather equivalently, think of $\lambda$ as the appropriate 't Hooft coupling \cite{tHooft:1973alw} for this theory).

The path integral for quantum mechanics in $N$ dimensions follows the same steps as for one-dimensional quantum mechanics, except that there is a quantum field for every dimension, so we end up with
\be
\label{z2}
Z=\int {\cal D}\vec{\phi} e^{-S_E}\,,\quad  S_E=\int_0^\beta \left[\frac{1}{2}\left(\partial_\tau \vec{\phi}\right)^2+\frac{\lambda}{N} \left(\vec{\phi}^2\right)^2\right]\,,
\ee
and $\vec{\phi}=\left(\phi_1,\phi_2,\ldots,\phi_N\right)$.

Instead of a hard path integral over a single field $\phi$ as in (\ref{z1}), we now have a path integral over multiple fields $\vec{\phi}$ which are all coupled together. If anything, this seems much harder than our original hard problem, so it doesn't look like we've made any progress here.

Don't despair yet, I have a trick down my sleeve!

The trick is that I can solve an integral over a Dirac $\delta$ function:
\be
\label{trick}
\int d\sigma \delta (\sigma-f) = 1\,,
\ee
for any real $f$. I can write a product of these integrals, and obtain a ``path-integral $\delta$'':
\be
\prod_\tau \int d\sigma(\tau) \delta\left(\sigma(\tau)-f(\tau)\right)=\int {\cal D}\sigma \delta\left(\sigma-f\right)=1\,.
\ee
Since this is true for any function $f(\tau)$ on the Euclidean circle, I can take $f(\tau)=\vec{\phi}^2(\tau)$ and thus re-write the partition function (\ref{z2}) as
\be
\label{z3}
Z=\int {\cal D}\vec{\phi}{\cal D}\sigma \delta(\sigma-\vec{\phi}^2) e^{-S_E}\,,\quad
S_E=\int_0^\beta d\tau \left[\frac{1}{2}\left(\partial_\tau \vec{\phi}\right)^2+\frac{\lambda}{N} \sigma^2\right]\,.
\ee
Having a delta function inside a path integral is un-field theorist, so I use
\be
\delta(x)=\int d\zeta e^{i \zeta x}\,,
\ee
to rewrite the path integral again as
\be
\label{z4}
Z=\int {\cal D}\vec{\phi}{\cal D}\sigma {\cal D}\zeta  e^{-S_E}\,,\quad
S_E=\int_0^\beta d\tau \left[\frac{1}{2}\left(\partial_\tau \vec{\phi}\right)^2+\frac{\lambda}{N} \sigma^2-i\zeta \left(\sigma-\vec{\phi}^2\right)\right]\,.
\ee

In this form, we have a path integral with two auxiliary fields $\sigma, \zeta$, but since the action for $\sigma$ is quadratic, we can integrate out $\sigma$ explicitly:
\be
\label{z5}
Z=\int {\cal D}\vec{\phi}{\cal D}\zeta  e^{-S_E}\,,\quad
S_E=\int_0^\beta d\tau \left[\frac{1}{2}\left(\partial_\tau \vec{\phi}\right)^2+i \zeta \vec{\phi}^2+\frac{N \zeta^2}{4 \lambda}\right]\,.
\ee

As a side remark,  rewriting of the path integral for quartic potential using an auxiliary field is known as a Hubbard-Stratonovic transformation in the literature. When I started working on this, I didn't know about Hubbard-Stratonovic, so I came up with this version which works for other potentials of the form $V(x)\propto x^\alpha$ as well, not just $\alpha=4$ \cite{Romatschke:2019rjk}. Apparently, sometimes ignorance is an advantage when working on a new subject.

The partition function (\ref{z5}) is quadratic in the field $\vec{\phi}$, so we can formally integrate out those fields as well, giving
\be
\label{z6}
Z=\int {\cal D}\zeta  e^{-S_E}\,,\quad
S_E=\frac{N}{2}{\rm Tr}\ln \left[-\partial_\tau^2+2 i \zeta\right]+\int_0^\beta d\tau \frac{N \zeta^2}{4 \lambda}\,.
\ee

So far, everything has been exact.

Splitting the auxiliary field $\zeta$ into zero-mode and fluctuations
\be
\label{split}
  \zeta(\tau)=\zeta_0+\zeta^\prime(\tau)\,,
  \ee
  we have
  \be
  S_E=\frac{N}{2}{\rm Tr}\ln \left[-\partial_\tau^2+2 i \zeta_0\right]+\frac{N \beta \zeta_0^2}{4 \lambda}+{\cal O}(\zeta^{\prime 2})\,.
  \ee
  The path integral over the fluctuations $\zeta^\prime$ cannot be calculated analytically in closed form. However, since it is a single field, the integral over the fluctuations cannot give a contribution of order $e^{{\cal O}(N)}$ to the path integral. So in the limit of large N, the (complicated) contribution from the fluctuations are sub-dominant.

  The calculation simplifies in the large N limit!

  For $N\gg 1$, we thus have
  \be
  \label{z7}
\lim_{N\gg 1}Z=\int d\zeta_0  e^{-S_{R0}}\,,\quad
S_{R0}=\frac{N}{2}{\rm Tr}\ln \left[-\partial_\tau^2+2 i \zeta_0\right]+\frac{N \beta \zeta_0^2}{4 \lambda}\,.
\ee

Instead of a path integral, the large N partition function is given in terms of a single integral, but the expression in the action still needs some work. (Aside: I've sneaked in the label ``R0'' for ``resummation level 0'' here, which is useful for the discussion in the following).

Since $\zeta_0$ is $\tau$-independent, it effective acts as a mass term, and we can calculate the trace of the logarithm of the operator as
\be
   {\rm Tr}\ln \left[-\partial_\tau^2+2 i \zeta_0\right]
   =\sum_n \langle n | \ln \left[-\partial_\tau^2+2 i \zeta_0\right] |n \rangle
   =\sum_n \ln \left[\omega_n^2+2 i \zeta_0\right]\,, 
   \ee
   when using $\langle \tau | n \rangle = e^{i \omega_n \tau}$ with $\omega_n=2 \pi n T$ the bosonic Matsubara frequencies. The ``thermal'' sum can be calculated using methods from thermal quantum field theory \cite{Laine:2016hma}, or by straightforward comparison to the partition function of the harmonic oscillator. Let's do the latter: for the harmonic oscillator, the partition function is
   \be
   \label{zo}
   Z_{HO}=\int {\cal D}\phi e^{-\frac{1}{2}\int_0^\beta \left[\dot\phi^2+m^2 \phi^2\right]}=e^{-\frac{1}{2} {\rm Tr}\ln \left[-\partial_\tau^2+m^2\right]}\,,
   \ee
   because it is a Gaussian integral.
   But we know the spectrum of the harmonic oscillator is $E_n=m \left(n+\frac{1}{2}\right)$, so we can calculate the harmonic oscillator partition function as
   \be
   Z_{HO}=\sum_{n=0}^\infty e^{-\beta E_n}=\frac{1}{2\sinh\left(\frac{m \beta}{2}\right)}\,.
   \ee
   Comparing the last two equation leads to
   \be
   \frac{1}{2}{\rm Tr}\ln \left[-\partial_\tau^2+m^2\right]=\ln\left[2\sinh\left(\frac{m \beta}{2}\right)\right]
   \ee

\begin{figure}
  \centering
  \includegraphics[width=0.7\linewidth]{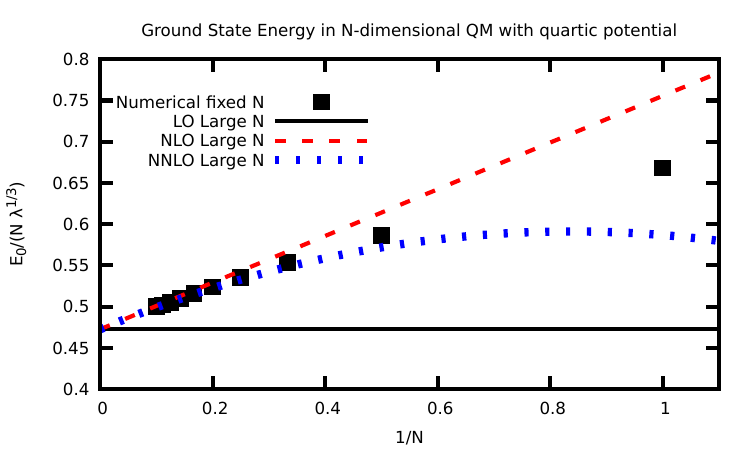}
  \caption{\label{fig:one} Ground state energy $\frac{E_0}{\lambda^{\frac{1}{3}} N}$ as a function of components $N$. Shown are numerical results from table \ref{tab:one} for $n=31$, and the analytic results in the large N limit: LO from (\ref{largeLO}), NLO from (\ref{largeNLO}, and NNLO is left as a homework problem. }
  \end{figure}
   
   As a consequence, we get for (\ref{z7})
 \be
  \label{z8}
\lim_{N\gg 1} Z= \int d\zeta_0  e^{-N \ln\left[2\sinh\left(\frac{\sqrt{2 i \zeta_0} \beta}{2}\right)\right]-\frac{N \beta \zeta_0^2}{4 \lambda}}\,.
\ee
This is the expression for the partition function of quantum mechanics in $N\gg 1$ dimensions at finite temperature. If we care about the ground state energy, we want to consider the low temperature limit $\beta\rightarrow \infty$. In this limit, the result simplifies to
 \be
  \label{z9}
\lim_{\beta \gg 1}\lim_{N\gg 1} Z= \int d\zeta_0  e^{-\frac{N \beta \sqrt{2 i \zeta_0}}{2}-\frac{N \beta \zeta_0^2}{4 \lambda}}\,.
\ee
For large $N$, the exponential is typically very small, except for the regions of the integral where the action is at a minimum. This is formally encoded in the saddle point method, so that integrals such as (\ref{z9}) can be evaluated exactly in closed form at large N. 
We find
 \be
  \label{z10}
  \lim_{\beta \gg 1}\lim_{N\gg 1} Z=e^{-\beta E(\zeta^*)}\,,
  \ee
  where $\zeta_0=\zeta^*$ is the solution to the saddle point condition
  \be
  \label{saddle1dQM}
  \frac{d E(\zeta^*)}{d\zeta^*}=0=\frac{N}{2\sqrt{2 i \zeta^*}}+\frac{N \zeta^*}{2\lambda} \longrightarrow i\zeta^*=\frac{(2\lambda)^{\frac{2}{3}}}{2}\,.
  \ee
  Plugging this saddle point back into the partition function, we get
  \be
  \label{largeLO}
  E(\zeta^*)=\frac{3 (2\lambda)^{\frac{1}{3}} N}{8}\simeq \lambda^{\frac{1}{3}} N\times 0.47247\ldots
  \ee
  Comparison between (\ref{z10}) and (\ref{z0}) shows that this is the ground state energy for quantum mechanics in $N\gg 1$ dimensions interacting via quartic potential. It is exact in the large N limit, and is smoothly connected to the ground state energy for finite, but large $N$, cf. Fig.~\ref{fig:one}. But even if we boldly extrapolate this result to $N=1$, we find that it only differs from the numerically calculated ground state energy of the one-dimensional quartic  anharmonic oscillator 
  \be
  \label{exactN1E0}
  E_0=\lambda^{\frac{1}{3}}\times 0.66799\ldots
  \ee
  by only about 30 percent (see Tab.\ref{tab:one} in the appendix and Ref.~\cite{Hioe:1978jj}).

We can add a $\frac{1}{N}$ improvement to the large N result of the ground state energy without too much trouble. Expanding $S_E$ in the exact partition function (\ref{z6}) to second order in fluctuations around the saddle: $\zeta=\zeta_0+\zeta^\prime(\tau)$ and performing a Fourier-transform 
\be
\zeta^\prime(\tau)=\int \frac{dk}{2\pi} e^{i k \tau} \zeta^\prime(k)\,.
\ee
In the zero temperature limit, we obtain
\be
\lim_{\beta \gg 1}\lim_{N\gg 1}Z=e^{-\frac{3 (2\lambda)^{\frac{1}{3}}}{8} N \beta}\int {\cal D}\zeta^\prime e^{-\int \frac{dk}{2\pi} \frac{N |\zeta^{\prime}(k)|^2}{4\lambda}-2 N \int \frac{dk}{2\pi} |\zeta^{\prime}(k)|^2\Pi(k)}\,,
\ee
with
\be
\label{piqm}
\Pi(k)=\frac{1}{2}\int \frac{dp}{2\pi}\frac{1}{(p^2+(2\lambda)^{\frac{1}{3}})((p+k)^2+(2\lambda)^{\frac{1}{3}})}=\frac{1}{2}\frac{1}{(2\lambda)^{\frac{1}{3}}(k^2+4 (2\lambda)^{\frac{2}{3}})}\,.
\ee
Performing the path integral over $\zeta^\prime$ leads to the large N ground state energy given by
\be
\label{largeNLO}
E_0=\frac{3 (2\lambda)^{\frac{1}{3}}}{8} N+\frac{1}{2}\int \frac{dk}{2\pi}\ln \left(1+\frac{2 (2\lambda)^{\frac{2}{3}}}{k^2+4 (2\lambda)^{\frac{2}{3}}}\right)=(2\lambda)^{\frac{1}{3}}\left(\frac{3 }{8} N+\frac{\sqrt{6}-2}{2}\right)+{\cal O}(N^{-1})\,.
\ee
Calculating the NNLO large N correction is possible with similar techniques, and obtaining the result (\ref{largeNNLO}) is left as an exercise (see below).

Extrapolating the NLO ground state energy for N=1 and comparing to the numerically calculated result for the N=1 theory (\ref{exactN1E0}), one finds that the NLO result is off by only about 13 percent.  Agreement with quantum mechanics in higher dimensions is better, as can be seen in Fig. ~\ref{fig:one}. Clearly, large N expansion techniques work quantitatively well in capturing the ground state energy for quantum mechanics at fixed and not too small N.

As a final note, let me point out that the fact that the NNLO correction does not improve on the disagreement for N=1, but helps with larger N, is in agreement with the expectation that the large N series expansion is asymptotic, just like the perturbative series expansion.

\subsection*{Guide to further reading}

Considering N-component field theory in dimension less than four is an interesting application of the above techniques. Here are a few suggestions for further reading

\begin{itemize}
\item
  The saddle point (\ref{saddle1dQM}) is not on the integration contour for $\zeta_0$, which are the real numbers. In order to access it, the integration contour needs to be deformed into the complex plane and becomes a ``thimble''. Ref.~\cite{Aarts:2013fpa} provides an excellent practitioner's introduction to the theory of Lefshetz thimbles.
\item
  Time-dependent quantum mechanics at large N, including the calculation of so-called Ljapunov exponents was studied in a series of papers in Refs.~\cite{Trunin:2021lwg,Kolganov:2022mpe,Trunin:2023xmw,Trunin:2023rwm}.
\item
  The scalar O(N) model in 2+1 dimensions was studied at finite temperature in \cite{Romatschke:2019ybu}. In this case, the field theory is super-renormalizable, and the large N expansion allows solution of this field theory for all values of the coupling. In particular, this includes a calculation of the exact large N shear viscosity coefficient \cite{Romatschke:2021imm}.
\item
  Theories with fermions, as well as certain supersymmetric theories in 2+1 dimensions can also be solved with the same technique, see \cite{DeWolfe:2019etx,Pinto:2020nip}.
\item
  Three dimensional QED with many flavors of electrons does not suffer from the problems encountered in four dimensions and can also be solved with similar techniques. The thermodynamics of large $N_f$ QED$_3$ was worked out in Ref.~\cite{Romatschke:2019qbx}, and the curious ``fractional photon'' in the strong coupling limit was pointed out in Ref.~\cite{Romatschke:2019mjm}. While it is possible to calculate transport coefficients in the strong coupling and large $N_f$ limit of QED$_{3}$/QCD$_{3}$ along the lines of Ref.~\cite{Moore:2001fga,Aarts:2005vc}, no such results currently exist in the literature. This is a typical example of an unclaimed patch in the Large N wonderland.
\item
  The O(N) model in 2+1 dimensions was conjectured to have a gravity dual in the strong coupling limit, cf. Ref.~\cite{Klebanov:2002ja}.
  There are encouraging works on reconstructing the bulk geometry from the boundary field theory in Refs.~\cite{Aharony:2020omh,Aoki:2016ohw}.
\item
  Higher dimensional O(N) models are not thought to be perturbatively renormalizable. However, O(N) models in odd dimensions (in particular in five dimensions) may be non-perturbatively renormalizable \cite{Parisi:1975im}. This has led to recent studies of O(N) models in odd dimensions, e.g. in Refs.~\cite{Li:2016wdp,Giombi:2019upv,Grable:2022swa}.
  \end{itemize}

\subsection*{Homework Problems Lecture 1}

\begin{enumerate}
\item[1.1]
Calculate $E_0$ in one-dimensional quantum mechanics with Hamiltonian (\ref{ham1}) using perturbation theory $\lambda \ll 1$. Compare your result to the numerically obtained result (\ref{exactN1E0}) and discuss.
\item[1.2]
  Calculate $E_0$ in N-dimensional quantum mechanics with Hamiltonian (\ref{ham2}) to order NNLO (including terms of order $N^{-1}$ in $E_0$) in a large N expansion. Show that
  \be
  \label{largeNNLO}
  E_0^{\rm NNLO}\simeq - 0.1689 N^{-1} \lambda^{\frac{1}{3}}\,.
  \ee
\item[1.3]
  Instead of quantum mechanics, now consider quantum field theory in 2+1 dimensions with Euclidean action
  \be
  \label{21d}
  S_E=\int_0^\beta d\tau \int d^2x \left[\frac{1}{2}\partial_\mu \vec{\phi} \cdot \partial_\mu \vec{\phi} + \frac{\lambda}{N}\left(\vec{\phi}^2\right)^2\right]\,.
  \ee
  Using the same techniques as for quantum mechanics, find the expression for the LO large N partition function $Z$ at finite temperature equivalent to (\ref{z8}). Defining the entropy density as $s=\frac{d}{d T} \frac{\ln Z}{\beta V}$, evaluate it at infinite coupling $s_{\infty}\equiv \lim_{\lambda \rightarrow \infty} s$. Show that
  \be
  \frac{s_{\rm \infty}}{s_{\rm free}}=\frac{4}{5}\,,
  \ee
  where $s_{\rm free}$ is the thermal entropy density of N free bosons in 2+1 dimensions.
\item[1.4]
  Consider again quantum field theory in 2+1 dimensions with Euclidean action (\ref{21d}). In Fourier space, the propagator for the scalar field $\phi$ at zero temperature can be parametrized as $G(k)=(k^2)^{-1+\frac{\eta}{2}}$ with $\eta$ the critical exponent. Calculate the first non-vanishing term of the critical exponent in a large N expansion and show that in the strong coupling limit $\lambda\rightarrow \infty$
  \be
  \eta=\frac{8}{3 N\pi^2}+{\cal O}\left(N^{-2}\right)\,.
  \ee

\end{enumerate}
\newpage
\section*{Lecture 2: Non-relativistic Neutrons}

\begin{figure}[t]
  \centering
  \includegraphics[width=0.7\linewidth]{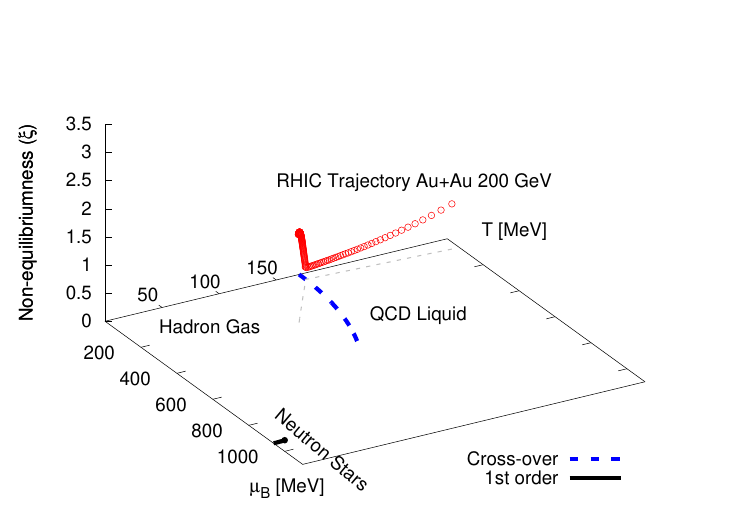}
  \caption{\label{fig:two} Sketch of what we know about the QCD phase diagram, adapted from Ref.~\cite{Romatschke:2016hle}. Axis are the equilibrium temperature $T$, baryon chemical potential $\mu_B$ and the parameter $\xi$ corresponding to deviations from equilibrium. Deconfinement cross-over and liquid-gas first order phase transitions are marked.  Areas relevant to neutron stars and relativistic heavy ion collisions -- such as gold ion collisions at center-of-mass energies of $\sqrt{s}=200$ GeV per nucleon pair at the Relativistic Heavy Ion Collider (RHIC) as well as their projection on the equilibrium $T,\mu_B$ plane (grey dashed lines) --  are indicated. See original reference for details.}
  \end{figure}

Consider the QCD phase diagram, sketched in Fig.~\ref{fig:two}. Most regions of this phase diagram are hard to access using first-principles QCD calculations, and this is especially true for the region of low temperature and finite baryon density relevant for neutron stars.

I only know of one exception to this statement: effective field theory (EFT).

EFTs are bona-fide field theories that are constructed out of the known symmetries, relevant degrees of freedoms, and a derivative expansion. Some well-known EFTs are chiral effective theory \cite{Gasser:1984gg} and relativistic fluid dynamics \cite{Baier:2007ix}.

EFTs have distinct advantages: they correspond to controlled, improvable first-principles calculations, and are often possible in regions where other approaches fail.

The main disadvantage to EFTs is that they invariably contain a finite number of free parameters that need to be fixed by other means, e.g. from experiment. 

In the following, I will consider a particular EFT for QCD at low temperature and finite baryon density relevant for neutron stars: pionless EFT, denoted as $\slashed{\pi}$ EFT \cite{Bedaque:2002mn}.

To build $\slashed{\pi}$ EFT, consider the energy scales relevant for low-temperature QCD: the nucleon masses $M\sim 940$ MeV, the pion masses $m_\pi \sim 135$ MeV and the deuteron binding energy $B\sim 2.2$ MeV. If we aim at a theory that only captures the deuteron, we need to include the nucleons, but can neglect excitations with energies much less than the pion mass. Hence we are driven to consider a theory of non-relativistic nucleons with kinetic energy $E_{\rm kin}\ll m_\pi$, so pions are not needed in this description, hence the name.

$\slashed{\pi}$ EFT for interacting nucleons has been fleshed out in a series of papers \cite{Kaplan:1998we,Chen:1999tn,Epelbaum:2008ga}, but for this lecture I want to focus on an even simpler version of $\slashed{\pi}$ EFT: pure neutron $\slashed{\pi}$ EFT. While inappropriate for describing nuclei such as the deuteron, this theory would be relevant for a very neutron rich environment. Can you think of one?

To build the EFT, we note that neutrons are fermions, and since we consider non-relativistic neutrons, we describe them as two-component spinors $\psi=\left(\begin{array}{c}\psi_\uparrow\\\psi_\downarrow\end{array}\right)$. Only neutrons, no anti-neutrons are included, because the energy scales for pair-production are much above the relevant scale of the theory. Free non-relativistic neutrons obey the Schr\"odinger equation, which can be turned into a Lagrangian density:
  \be
  {\cal L}=\psi^\dagger \left(i\partial_t +\frac{\vec{\nabla}^2}{2 M}\right)\psi\,.
  \ee

Field theorists accustomed to relativistic fields will find that this form also arises from taking the non-relativistic limit of the free Dirac fermion Lagrange density $\bar\Psi i \slashed{\partial} \Psi$.

The above Lagrangian describes free (non-interacting) non-relativistic neutrons. This is boring. In order to have something of interest, we need to include interactions. In an EFT, one writes down all possible interactions allowed by symmetry, such as two-neutron, three-neutron, four-neutron, etc. interactions. All of these come with unknown coefficients that need to be fixed by other means, e.g. experiment. However, the lowest-order interaction is that of a two-neutron singlet ``contact term'' (no derivatives), such that \cite{Bedaque:2002mn}
\be
{\cal L}_I=-\frac{C_0}{4}\left(\psi \sigma_y \psi\right)^\dagger \left(\psi \sigma_y \psi\right)\,,
\ee
where $\sigma_y=\left(\begin{array}{cc} 0 & -i \\
  i & 0\end{array}\right)$ is the second Pauli matrix. As promised, $C_0$ is a coefficient that needs to be fixed by other means. In the present case, this can be done by calculating the scattering amplitude and comparing to the corresponding scattering amplitude resulting from solving the Schr\"odinger equation (see appendix \ref{sec:efts} for the explicit matching in the case of bosons). One finds
  \be
  \label{c0def}
  C_0=\frac{4 \pi a_0}{M}\,,
  \ee
  where $a_0$ is the s-wave scattering length for neutrons. Fortunately, the s-wave scattering lengths for neutrons is well known experimentally \cite{PhysRevLett.83.3788} as
  \be
  \label{a0}
  a_0\simeq -18.5 \ {\rm fm}\,,
  \ee
  which together with the known nucleon mass $M$ fixes the parameters of the theory. We are now ready to calculate!

  Let's jump right in and write down the grand-canonical partition function for pure-neutron $\slashed{\pi}$ EFT with spin-singlet interaction:
  \be
  \label{zSP}
  Z=\int {\cal D}\psi e^{-S_E+(\mu_B-M) N}\,,
  \ee
  where
  \be
  S_E=\int_0^\beta d\tau \int d^3x \left[\psi^\dagger \left(\partial_\tau-\frac{\vec{\nabla}^2}{2m}\right)\psi+\frac{C_0}{4}\left(\psi \sigma_y \psi\right)^\dagger \left(\psi \sigma_y \psi\right)\right]\,,
  \ee
  is the Euclidean action corresponding to analytically continuing the Lagrangian density ${\cal L}$ above to Euclidean time $\tau \in [0,\beta]$, $\mu_B$ is the baryon chemical potential, and
  \be
  N=\int_0^\beta d\tau \int d^3x \psi^\dagger \psi\,,
  \ee
  is the neutron number. Since the baryon chemical potential only appears in the combination $\mu_B-M$, it is useful to denote this ``excess'' chemical potential as
  \be
  \mu\equiv \mu_B-M\,.
  \ee

  With the theory defined by the grand-canonical partition function (\ref{zSP}), obtaining observables such as the pressure $p\equiv \frac{\ln Z}{\beta V}$, the baryon density $n\equiv \frac{\partial}{\partial \mu}p$ and the excess energy density (equal to energy density minus nucleon rest mass) $\epsilon=\mu n-p$ is `''just'' a matter of solving the many-body partition function.

  However, even for this admittedly simple EFT, exact solutions for $Z$ are hard because of the 4-fermi interaction term in (\ref{zSP}):
  \be
  \left(\psi \sigma_y \psi\right)^\dagger \left(\psi \sigma_y \psi\right) = 4 \left(\psi_\downarrow \psi_\uparrow\right)^\dagger \left(\psi_\downarrow \psi_\uparrow\right)\,.
  \ee

  But we learned in lecture 1 how to deal with such quartic interactions in a large N framework! Let's make use of this knowledge!

  Instead of a single neutron species, consider $N$ neutron species $\psi\rightarrow \psi_f=\left(\psi_1,\psi_2,\ldots,\psi_N\right)$. You may think of these either as fictitious extra particles, or for $N=2$, as a very crude way of including the proton into the description. In either case, we will use $\frac{1}{N}\ll 1$ as a small expansion parameter unrelated to any other parameter in the theory, which allows us to perform non-perturbative calculations of the theory.

  In complete analogy to the case of quantum mechanics studied in lecture 1, we generalize the interaction term to the N-component case as
  \be
  C_0 \left(\psi_\downarrow \psi_\uparrow\right)^\dagger \left(\psi_\downarrow \psi_\uparrow\right)\rightarrow \frac{C_0}{N}\left(\psi_{\downarrow,f} \psi_{\uparrow,f}\right)^\dagger \left(\psi_{\downarrow,g} \psi_{\uparrow,g}\right)\,,
  \ee
  where the ``flavor'' indices $f,g$ run from 1 to N and Einstein sum convention is used to suppress the summation symbols.

  Next, introduce the complex auxiliary field $\zeta$ through inserting the identity
  \be
  1=\int {\cal D}\zeta e^{N \int_x \frac{\zeta^* \zeta}{C_0}}
  \ee
  (note that this makes sense because $C_0\propto a_0$ is negative for neutrons, cf. (\ref{a0})). Now shifting
  \be
  \zeta\rightarrow \zeta-\frac{i C_0}{N}\psi_{\downarrow, f}\psi_{\uparrow, f}
  \ee
  then leads to the auxiliary-field formulation for N-component pure-neutron $\slashed{\pi}$ EFT:
  \be
  Z=\int {\cal D}\psi {\cal D}\zeta e^{-\int_x\left[\psi^\dagger_{f}\left(\partial_\tau-\frac{\vec{\nabla}^2}{2M}-\mu\right) \psi_f + i \zeta^*\psi_{\downarrow,f}\psi_{\uparrow, f}-i \zeta \psi^\dagger_{\uparrow,f}\psi^\dagger_{\downarrow,f}-\frac{N \zeta \zeta^*}{C_0}\right]}\,.
  \ee
  In this form, all the fermions enter as bilinears into the path integral action. They can be compactly brought into the form
  \be
  \Psi^\dagger_f G^{-1} \Psi_f\,,
  \ee
  with the two-component composite (Nambu-Gorkov) spinor
  \be
  \Psi=\left(\begin{array}{c}
    \psi_\uparrow\\
    \psi^\dagger_\downarrow
    \end{array}\right)\,,
  \ee
  and the inverse propagator in matrix form
  \be
  G^{-1}=\left(\begin{array}{cc}
    \partial_\tau -\frac{\vec{\nabla}^2}{2M}-\mu & - i \zeta\\
    i \zeta^* & \partial_\tau+\frac{\vec{\nabla}^2}{2M}+\mu
    \end{array}\right)\,.
  \ee
  Since the fermions enter the action quadratically, they can be integrated out:
  \be
  Z=\int {\cal D}\zeta e^{N \ln \rm det G^{-1}+\frac{N}{C_0}\int_x \zeta^* \zeta}\,.
  \ee

  So far, everything has been exact. However, in the large N limit, the remaining path integral simplifies considerably because of the same reason outlined in quantum mechanics after Eq.~(\ref{split}): the leading large N saddle corresponds to constant $\zeta$, or equivalently the zero mode $\zeta_0$. In the literature, it is customary to denote $i \zeta_0^*\equiv \Delta$, and (with hindsight) assume $\Delta$ to be real. Then the large N partition function becomes
  \be
  \lim_{N\gg 1} Z=\int d\Delta e^{N \beta V p(T,\Delta)}\,,
  \ee
  with
  \be
  \label{p1}
  p(T,\Delta)=\frac{\Delta^2}{C_0}+T \sum_n \int \frac{d^3{\bf k}}{(2\pi)^3} \ln \left[\tilde \omega_n^2+(\epsilon_k-\mu)^2+\Delta^2\right]\,,
  \ee
  where $\tilde \omega_n=\pi T \left(2n+1\right)$ the fermionic Matsubara frequencies and $\epsilon_k=\frac{{\bf k}^2}{2M}$ the non-relativistic kinetic energy.

  In the zero temperature limit, the thermal sum in (\ref{p1}) becomes an integral which is straightforward to solve:
   \be
  \label{p2}
  p(0,\Delta)=\frac{\Delta^2}{C_0}+\int\frac{d^3{\bf k}}{(2\pi)^3} \sqrt{(\epsilon_k-\mu)^2+\Delta^2}\,.
  \ee
  The remaining integral over momenta ${\bf k}$ can likewise be calculated in closed form when using dimensional regularization. (It is of course also possible to use old-fashioned cut-off regularization, but why use an old combustion engine when you can drive an electric car instead?) Expanding the square root and using the identities from Ref.~\cite{Nishida:2006eu}, one finds \cite{Lawrence:2022vwa}
  \be
  \label{p3}
  p(0,\Delta)=\frac{\Delta^2}{C_0}+\frac{2\mu }{5} \frac{(2 M \mu)^{\frac{3}{2}}}{3\pi^2} g\left(\frac{\mu}{\sqrt{\mu^2+\Delta^2}}\right)\,,
    \ee
    where the function $g(y)=y^{-\frac{5}{2}}\left[(4 y^2-3)E\left(\frac{1+y}{2}\right)+\frac{3+y-4 y^2}{2} K\left(\frac{1+y}{2}\right)\right]$ is expressed using $E,K$, the complete elliptic integrals of the first and second kind, respectively.

    To leading order in the large N and low temperature limit, the grand-canonical path integral is then given as
    \be
    \lim_{\beta\gg 1}\lim_{N\gg 1} Z=e^{N \beta V p(0,\Delta)}\,,
    \ee
    with $\Delta$ being the solution of the saddle point condition
    \be
    \label{saddleN}
    0 = \frac{dp(0,\Delta)}{d\Delta}\,.
    \ee

    We have a solution!

    Now let's see if the solution is any good. We need the neutron density, which we can calculate as
    \be
    \label{nnnLO}
    n=\frac{d p(0,\Delta)}{d\mu}=\frac{(2 M \mu)^{\frac{3}{2}}}{3\pi^2} g(y)\,,
    \ee
    where $y=\frac{\mu}{\sqrt{\mu^2+\Delta^2}}$ and we have used the saddle point condition (\ref{saddleN}) to simplify the expression. Using $n$ and the zero-temperature pressure $p(0,\Delta)$ we can construct the energy density $\epsilon$, and in particular the energy per particle
    \be
    \label{EoSLO}
    \frac{E}{N}=\frac{\epsilon}{n}=\mu \left(\frac{3}{5}+\frac{3 \pi \Delta^2}{8 \mu^2 g(y) \sqrt{2 M \mu a_0^2} }\right)\,.
    \ee

\begin{figure}
  \centering
  \includegraphics[width=0.7\linewidth]{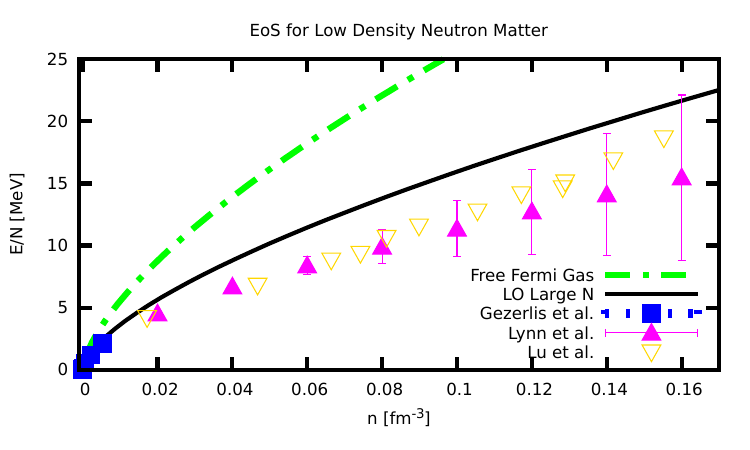}
  \caption{\label{fig:three} Energy per particle for pure neutron matter as a function of density. Shown are results for the free Fermi gas $\frac{E}{N}=\frac{3\mu}{5}$, the LO large N result (\ref{EoSLO}) and Monte Carlo results from three different groups: Gezerlis et al.~\cite{Gezerlis:2009iw,Gezerlis:2011gq}, Lynn et al. ~\cite{Lynn:2015jua,Tews:2018kmu} and Lu et al. ~\cite{Lu:2018bat}.}
  \end{figure}
    
    For a given value of $\mu$, we can numerically calculate the value of $\Delta$ from solving the saddle point condition (\ref{saddleN}). With $\mu,\Delta$, we can then calculate $n$ and $\frac{E}{N}$. How do our leading order large N results compare to other methods?

    The relevant comparison is shown in Fig.~\ref{fig:three}, where the lO large N result for the energy per particle for pure neutron matter is compared to the results from three other groups. One finds that the LO large N result for $\frac{E}{N}$ is about 30 percent higher than the considerably more complex calculations from Refs.~\cite{Gezerlis:2009iw,Gezerlis:2011gq,Lynn:2015jua,Tews:2018kmu,Lu:2018bat}.

The 30 percent difference is surprisingly similar to what we found when comparing the LO large N result to the N=1 ground state energy for the quartic oscillator in quantum mechanics in the first lecture. In that lecture, we found that going to NLO in the large N expansion was straightforward, and just involved a Gaussian integral, yet reduced the difference with the N=1 value by a factor of two.

Not surprisingly, calculating the NLO large N correction to the grand-canonical partition function can be done with similar ease here \cite{veillette2007large}. What is surprising, though, is that the equivalent NLO large N result for Fig.~\ref{fig:three} is not available in the literature! 

Another unclaimed patch in the Large N wonderland. Maybe you can help?

\subsection*{Guide to further reading}

\begin{itemize}
\item
  The coupling $C_0$ in (\ref{c0def}) explicitly depends on the value of the s-wave scattering length $a_0$, which for certain atomic systems can be experimentally tuned by varying the external magnetic field $B$. Depending on the value of $B$, $a_0$ increases from zero to almost infinity at the so-called Feshbach resonance, wraps around and starts at almost negative infinity, and approaches zero again for larger values of $B$ \cite{gurarie2007resonantly}. Despite this extreme behavior of the coupling near the Feshbach resonance, experimental observables in these atomic experiments remain perfectly well-behaved. 
\item
  Transport coefficients can be calculated for the pure neutron matter theory in the large N limit for any coupling/density. Currently, only the LO large N result for so-called thermodynamic transport coefficients are known \cite{Shukla:2019shf,Lawrence:2022vwa}, but calculating shear viscosity along the lines of Ref.~\cite{Moore:2001fga,Aarts:2005vc} is doable.
\item
  Calculating the zero temperature limit of the grand canonical partition function to NLO in the large N limit exhibits a concrete example of non-commutative limits that was uncovered in Ref.~\cite{Gorda:2022yex}.
\end{itemize}

\subsection*{Homework Problems Lecture 2}

\begin{enumerate}
\item[2.1]
  In the literature, the strong coupling limit $a_0\rightarrow -\infty$ near a Feshbach resonance is called the ``Unitary Fermi Gas'' limit, whereas the weak coupling limit $a_0\rightarrow 0$ is called the ``Free Fermi Gas'' limit. Calculate the large N ``superfluid gap'' $\Delta$ from solving (\ref{saddleN}) in both of these limits and show that
  \be
  \lim_{a_0\rightarrow -\infty} \Delta \simeq 1.1622 \times \mu\,,\quad
  \lim_{a_0\rightarrow 0} \Delta \simeq e^{-\frac{\pi}{\sqrt{8 M \mu a_0^2}}-2+3 \ln 2}\times \mu\,.
  \ee
\item[2.2]
  In the Unitary Fermi Gas limit, the energy density can be expressed as
  \be
  \lim_{a_0\rightarrow -\infty} \epsilon = \frac{3}{5} n^{\frac{5}{3}} \frac{(3\pi^2)^{\frac{2}{3}}}{2 M}\times \xi\,,
  \ee
  with $\xi$ a pure number (the ``Bertsch parameter''). Calculate $\xi$ in the large N approximation and show that
  \be
  \lim_{N\gg 1}\xi\simeq 0.59\,.
  \ee
  \end{enumerate}

\newpage
\section*{Lecture 3: Negative Coupling and Triviality}

In lecture 1, we considered large N techniques for N-dimensional quantum mechanics, and found that the large N calculations gave improvable and reasonably accurate results for finite N, including down to N=1.

In lecture 2, we considered large N techniques for a four-dimensional (non-relativistic) quantum field theory of interacting neutrons, and we found that also here large N gave reasonable results even for N=1.

There are plenty of other examples I could cite about successes of large N calculations applied to observables at finite (and sometimes quite small) N.

It seems the method is sound and the math is trustworthy.

So how about we trust the math, even if its implications are non-intuitive?

Let's see where this ``trust the math'' axiom leads in the case of four-dimensional scalar field theory.

To be concrete, let's consider N-component scalars $\vec{\phi}=\left(\phi_1,\phi_2,\ldots, \phi_N\right)$ interacting via a quartic coupling with Euclidean action
\be
\label{4don}
S_E=\int d^4x \left[\frac{1}{2}\partial_\mu \vec{\phi}\cdot \partial_\mu \vec{\phi}+\frac{\lambda}{N}\left(\vec{\phi}^2\right)^2\right]\,.
\ee
This theory referred to as the O(N) model in the literature.

If you want to have a concrete physical system in mind, consider the Standard Model Higgs field is a two-component complex scalar $\Phi=\left(\begin{array}{c}
  \phi_1+i \phi_2\\
  \phi_3+i \phi_4\end{array}\right)$, which is equivalent to considering the O(N) model for $N=4$. Since $N=4$ is not that small, we might even expect our large N techniques to be quantitatively better in describing the Higgs sector than for instance the pure neutron case in lecture 2.

  The Euclidean action then defines the partition function for the theory in terms of a path integral $Z=\int {\cal D}\vec{\phi} e^{-S_E}$. Using exactly the same steps as in lecture 1, we can introduce an auxiliary field $\zeta$ to make the action quadratic in the field $\vec{\phi}$, so the path integral over $\vec{\phi}$ can be done in closed form:
  \be
  \label{L3z0}
  Z=\int{\cal D}\vec{\phi}{\cal D}\zeta e^{-\int_x \frac{1}{2}\vec\phi\left[-\partial_\mu \partial_\mu+2 i \zeta\right]\vec{\phi}-\frac{N}{4\lambda}\int d^4x \zeta^2}=\int {\cal D}\zeta e^{-\frac{N}{2}{\rm Tr}\ln\left[-\partial_\mu\partial_\mu+2 i \zeta\right]-\frac{N}{4\lambda}\int d^4x \zeta^2}\,.
  \ee

  Also, again just as in the case of quantum mechanics, when splitting the auxiliary field into a global zero mode $\zeta_0$ and fluctuations $\zeta^\prime$, the path integral over fluctuations does not contribute to the LO large N partition function, hence
  \be
  \label{L3z1}
  \lim_{N\gg 1}Z = \int d\zeta_0 e^{-\frac{N}{2}{\rm Tr}\ln\left[-\partial_\mu\partial_\mu+2 i \zeta_0\right]-\frac{N}{4\lambda}\int d^4x \zeta_0^2},
  \ee
where the quantum field theory partition function is now given in terms of a single integral (and not a path integral!).

Because $\zeta_0$ does not depend on position, it is a constant as far as the operator $\left[-\partial_\mu\partial_\mu+2 i \zeta_0\right]$ is concerned. Hence we can treat $2 i \zeta_0=m^2$ as a constant mass term and directly evaluate the trace of the operator, e.g. via dimensional regularization \cite[Eq.~2.72]{Laine:2016hma}
\be
\label{id0}
\frac{1}{2 {\rm vol}}{\rm Tr}\ln\left[-\partial_\mu\partial_\mu+m^2\right]=\frac{1}{2}\int \frac{d^{4-2\varepsilon}k}{(2\pi)^{4-2 \varepsilon}} \ln \left[k^2+m^2\right]=
-\frac{m^4}{64\pi^2}\left(\frac{1}{\varepsilon}+\ln \frac{\bar\mu^2 e^{\frac{3}{2}}}{m^2}\right)\,,
\ee
where ${\rm vol}=\int d^4x$ denotes the spacetime volume and $\bar\mu$ is the $\overline{\rm MS}$ renormalization scale. (For those unafraid of needlessly breaking Lorentz invariance, one can also do this calculation in  cut-off regularization, see homework problem 3.1 below).

The large N partition function then is given by
\be
  \label{L3z2}
  \lim_{N\gg 1}Z = \int d\zeta_0 e^{-{\rm vol}\times \frac{N\zeta_0^2}{4} \left[\frac{1}{\lambda}+\frac{1}{4 \pi^2 \varepsilon}+\frac{1}{4\pi^2}\ln \frac{\bar\mu^2 e^{\frac{3}{2}}}{2 i \zeta_0}\right]}.
  \ee
  After regularization, the expression for the partition function still has an uncanceled UV divergence for $\varepsilon\rightarrow 0$. This divergence can be canceled by introducing a suitable coupling-constant counterterm to the bare coupling $\lambda$ in a renormalization procedure. For the case at hand, we can non-perturbatively renormalize the theory by introducing the renormalized (running) coupling $\lambda_R$ as
  \be
  \label{renormalization}
  \frac{1}{\lambda}+\frac{1}{4 \pi^2 \varepsilon}\equiv \frac{1}{\lambda_R(\bar\mu)}\,.
  \ee

  Note that this renormalization procedure is non-perturbative because $\lambda$ contains an infinite number of terms with powers of $\lambda_R$. Also note that this renormalization procedure does not recover the LO perturbative renormalization when expanded in powers of the coupling, because the LO large N theory does not contain the full LO perturbative contribution (actually only one third of it, whereas the remaining 2/3 originate at NLO in the large N limit, cf. R1/R2 level resummation in Ref.~\cite{Romatschke:2019wxc}).

  Given the renormalization (\ref{renormalization}), one obtains the running coupling as
  \be
  \label{runningL}
  \lambda_R(\bar\mu)=\frac{4\pi^2}{\ln \frac{\Lambda_{\overline{\rm MS}}^2}{\bar\mu^2}}\,,
  \ee
  where $\Lambda_{\overline{\rm MS}}$  is an emergent scale of the theory. It is defined by the value of the scale $\bar\mu$ at which $\lambda_R$ diverges, e.g.
  \be
  \label{LP}
  \lambda_R(\Lambda_{\overline{\rm MS}})=\infty\,.
  \ee
  The scale $\Lambda_{\overline{\rm MS}}$ is commonly referred to as the ``Landau pole'' of the theory, even though it is clear from (\ref{runningL}) that $\lambda_R$ does not have a pole, but rather a logarithmic singularity at $\bar\mu=\Lambda_{\overline{\rm MS}}$.

\begin{figure}
  \centering
  \includegraphics[width=0.7\linewidth]{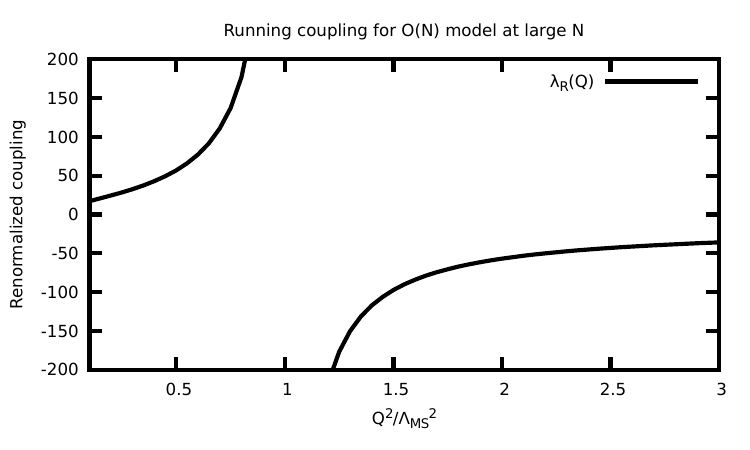}
  \caption{\label{fig:four} Exact large N running coupling $\lambda_R(Q)$ from Eq.~(\ref{runningL}). Figure from Ref.~\cite{Romatschke:2023sce}. See text for details.}
  \end{figure}
  
A plot of the running coupling is shown in Fig.~\ref{fig:four}. In particular, note that in the UV limit, the running coupling approaches zero from below:
\be
\label{negative}
\lim_{\bar\mu\rightarrow \infty}\lambda_R(\bar\mu)=0^-\,.
\ee
I cannot help point out the similarity of Fig.~\ref{fig:four} to the behavior of the coupling $C_0$ in (\ref{c0def}) in cold atom experiments as a function of the applied magnetic field, see e.g. Ref.~\cite{gurarie2007resonantly}.

  It is straightforward to calculate the $\beta$ function for this theory as
  \be
  \label{betafunc}
  \beta\equiv \frac{d \lambda_R(\bar\mu)}{d\ln \bar\mu^2}=\frac{4 \pi^2}{\ln^2 \frac{\Lambda_{\overline{\rm MS}}^2}{\bar\mu^2}}=\frac{\lambda_R^2(\bar\mu)}{4\pi^2} \geq 0\quad \forall\ \lambda_R \in \mathbb{R}\,.
  \ee
  Obviously, the $\beta$ function is positive, consistent with an ever-increasing running coupling, cf. Fig.~\ref{fig:four}.

  Before trying to make sense of these results, let me stress that the running coupling (\ref{runningL}), its negative value in the UV (\ref{negative}, the Landau pole (\ref{LP}) and the $\beta$ function (\ref{betafunc}) are exact results in the large N limit. In particular, their validity is not limited to a weak coupling domain, because we did not use a weak coupling expansion in obtaining them.

  Let's review the prevailing interpretation for these findings first, before heeding my advice of ``trust the math, even if it's non-intuitive''.

  By far the majority opinion of theoretical physicists is that a negative coupling, a positive $\beta$ function and/or a Landau pole are all fatal flaws of a continuum interacting quantum field theory.   Reviewing these one-by-one, it is possible to understand how the verdict ``fatal'' arises in each case. However, in the interest of keeping the lecture to its allotted time frame, I relegate this to the guide to further reading at the end.

  For now, let's ignore ``fatal flaw'' majority opinion, trust the math, and see where it leads us.

  So instead of giving up, we can ask the question: is there actually something wrong with the theory?

  In order to answer this question, we better calculate observables, so let's do that.

  The first observable we can look at is the mass of the field $\vec{\phi}$, which for $N=4$ would be nothing else but the Higgs boson mass. The large N Euclidean Green's function for $\vec{\phi}$ is given by $[-\partial_\mu\partial_\mu +2 i \zeta_0]^{-1}$, so at large N, the vector mass is determined through
  \be
  m^2=2 i \zeta_0\,,
  \ee
  where $\zeta_0$ is the location of the saddle point. After renormalization, the large N partition function (\ref{L3z2}) is given by
  \be
  \label{L3z3}
   \lim_{N\gg 1}Z = \int d\zeta_0 e^{-{\rm vol}\times \frac{N\zeta_0^2}{4} \left[\frac{1}{4\pi^2}\ln \frac{\Lambda_{\overline{\rm MS}}^2 e^{\frac{3}{2}}}{2 i \zeta_0}\right]},
   \ee
   from which the saddle point condition becomes\footnote{As an aside, note that any physical observable ${\cal O}$ must be renormalization-scale independent, $\frac{d {\cal O}}{d\bar\mu}=0$. It is gratifying to find that both the large N partition function (\ref{L3z3}) and the saddle point condition for the vector mass are explicitly renormalization scale independent.}
   \be
   \label{saddleL3}
   \frac{\zeta_0}{8\pi^2}\ln \frac{\Lambda_{\overline{\rm MS}}^2 e^{1}}{2 i \zeta_0} =0\,.
   \ee
   This saddle point condition implies two solutions for the vector mass squared:
   \be
   \label{twosolutions}
   m^2=0\,,\quad m^2=e \Lambda_{\overline{\rm MS}}^2\,.
   \ee
   The first of these corresponds to a vanishing vector mass expectation value, which corresponds to the prevailing assumption for the perturbative vacuum for the theory defined by (\ref{4don}). In the perturbative setup of the Electroweak sector of the Standard Model, one introduces a ``negative mass squared'' term $-m^2 \vec{\phi}^2$ (a tachyon) into the action in order to get spontaneous symmetry breaking, and one obtains a non-vanishing vector mass only after this construction. (Side remark: it does not strike me as particularly natural to set up a theory as a perturbation around a tachyonic vacuum, but that is the current prevailing physics setup for the Electroweak sector.)

   By contrast, the second solution (\ref{twosolutions}) corresponds to a non-perturbative vacuum where the vector mass is non-vanishing even though O(N) symmetry remains unbroken. This is clearly different from the Standard Model, already because the mass does not get put in ``by hand'' through the addition of a tachyon to the theory. In this situation, the Higgs mass becomes a prediction of the theory, not a parameter.

   But which of the two solutions (\ref{twosolutions}) is the right one?

   There is an easy way to decide this question, and hinges on calculating a second observable, the free energy $F$ of the theory. Namely, each of the two solutions will lead to a different value of the large N partition function, and hence the large N free energy. The correct solution to (\ref{twosolutions}) then is the one that has the lower free energy.

   Let's calculate: in the two cases, we get for the large N free energy 
   \be
   F_{m^2=0}=0\,,\quad F_{m^2=e \Lambda_{\overline{\rm MS}}^2}=-{\rm vol}\times \frac{N e^{2}\Lambda_{\overline{\rm MS}}^4}{128 \pi^2}\,.
   \ee
   Clearly, the non-perturbative solution has the lower free energy, and hence the perturbative vacuum must be unstable. It seems as if this actually agrees with the consensus opinion for the Standard Model Electroweak Sector as of September 2023.

   We thus find for the two observables (vector mass and free energy density) in the O(N) model:
   \be
   \label{dominant}
   m=\sqrt{e} \Lambda_{\overline{\rm MS}}\,,\quad \frac{F}{\rm vol}=\frac{N e^{2}\Lambda_{\overline{\rm MS}}^4}{128 \pi^2}\,.
   \ee
   Both of these are finite, non-vanishing and renormalization scale independent, despite the decidedly weird properties of the theory (\ref{negative}), (\ref{LP}), (\ref{betafunc}). Even better, they are parameter-free predictions for the Higgs mass and Higgs free energy in the case of N=4!

  How's that for a theory that doesn't exist/is trivial/is fatally flawed?

   Maybe trusting the math is not such a crazy suggestion after all.

Or could it be that the ``fatal flaw'' reveals itself only when we look at scattering?

\begin{figure}
  \centering
  \includegraphics[width=0.7\linewidth]{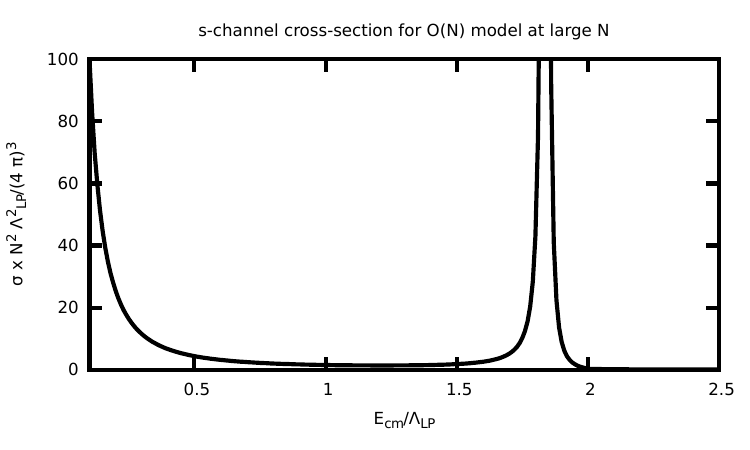}
  \caption{\label{fig:five} s-channel cross section for scattering in the 4d O(N) model to LO in large N, reproduced from Ref.~\cite{Romatschke:2022jqg}. }
  \end{figure}
   
   So let's calculate scattering cross-sections at large N. To this end, consider the connected, amputated four-point function
   \be
   {\cal M}=-\langle \phi_a(x_1) \phi_b(x_2) \phi_c(x_3) \phi_d(x_4)\rangle_{\rm conn.,amp.}
   \ee
   at large N. From (\ref{L3z0}), this becomes for the s-channel amplitude in momentum space
   \be
   {\cal M}(k)=D(k)\,,
   \ee
   where $D(x-y)=\langle \zeta(x) \zeta(y)\rangle$ is the auxiliary field propagator. The auxiliary field propagator can be calculated by again integrating out the vector field $\vec{\phi}$, and then expanding the action to second order in the fluctuation field $\zeta^\prime$. In complete analogy to Eq.~(\ref{piqm}), one finds
   \be
   \label{dk3}
   D(k)=\frac{1}{\frac{N}{8\lambda}+N \Pi(k)}\,,\quad
   \Pi(k)=\frac{1}{2}\int \frac{d^{4-2\varepsilon}p}{(2\pi)^{4-2 \varepsilon}} \frac{1}{p^2+m^2} \frac{1}{(p+k)^2+m^2}\,,
   \ee
   where $m$ is the large N vector mass for the dominant saddle (\ref{dominant}). 

   Eq.~(\ref{dk3}) contains the complete contribution to order $\frac{1}{N}$, but is fully non-perturbative in the coupling. To see this, note that (\ref{dk3}) can formally be expanded out in a power series in $\lambda$, obtaining
   \be
   \label{dk23}
   D(k)=\frac{8\lambda}{N}\sum_{n=0}^\infty\left(-8 \lambda \Pi(k)\right)^n=\vcenter{\includegraphics[width=0.5\linewidth]{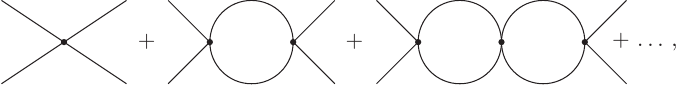}}\,.
   \ee
   where ``bubbles'' correspond to $\Pi(k)$, and each ``vertex'' corresponds to a factor of $\lambda$. The key lesson here is that when replacing $\lambda$ by (\ref{runningL}), then each term in (\ref{dk23}) contains a divergence at the Landau pole $\bar\mu=\Lambda_{\overline{\rm MS}}$, and in fact divergencies get worse at each order in perturbation theory. Naively, one could conclude that the theory is sick, and in fact many people have come to this conclusion.

   However, this is a breakdown of the perturbative expansion, and not the theory itself. This can easily be seen by noting that $D(k)$ can be evaluated in closed form. The momentum integral for $\Pi(k)$ is done in dimensional regularization finding a the UV divergence for $\Pi(k)$ for $\varepsilon\rightarrow 0$. Now this UV divergence for $\Pi(k)$ exactly cancels the UV-divergence from the coupling constant $\lambda$ when using the renormalization condition (\ref{renormalization}). One finds \cite{Romatschke:2022jqg}
   \be
   \label{auxiliary}
   D^{-1}(k)=\frac{N}{32\pi^2}\left[\ln \frac{\Lambda_{\overline{\rm MS}}^2 e^2}{m^2}-2\sqrt{1+\frac{4 m^2}{k^2}} {\rm atanh}\sqrt{\frac{k^2}{k^2+4m^2}}\right]\,.
   \ee
   The result is finite and well-behaved for all energies, even though its perturbative expansion (\ref{dk23}) was ripe with divergencies. The upshot is that the scattering amplitude in the O(N) model is inherently non-perturbative, but divergence-free, in the large N limit.

   To obtain the s-channel scattering amplitude, we need to analytically continuing $D(k)$ to Minkowski space as $k^2\rightarrow -E^2+{\bf k}^2 -{\rm sgn}(E)  i 0^+$. A plot of the s-channel cross section is shown in Fig.~\ref{fig:five}. Note again the explicit independence of $\sigma$ from the renormalization scale $\bar\mu$, as expected for a physical observable.

No pathologies are observed for scattering in the LO large N limit. The only curious finding is the presence of a stable bound state with a mass of $m_2\simeq 1.84 m$. 

Where are all the scary pathologies hiding?

I do not know....

  \subsection*{Guide to further reading}
  \begin{itemize}
  \item
    Obtaining a non-vanishing Higgs mass without introducing a negative mass squared term into the theory was considered a long time ago by Coleman and Weinberg in a famous paper on radiative corrections \cite{Coleman:1973jx}. The prediction for the Higgs mass in the so-called Coleman-Weinberg mechanism came out wrong, but that may be partly a consequence of doing the calculation perturbatively and throwing away terms ``not under perturbative control''.
  \item
    For many people, the Landau pole is a showstopper because perturbation theory breaks down, which on the other hand is not an issue if using techniques not limited to weak coupling (such as large N). Other people co-mingle the Landau pole with Landau's ghost, a tachyonic excitation that appears in perturbative QED. However, as discussed in Ref.~\cite{Romatschke:2022llf}, the large N O(N) model in four dimensions does not have a Landau ghost (even though it has a Landau pole), in contradistinction to perturbative QED.
    \item
    The original studies of the O(N) model in four dimensions date back to the 1970s \cite{Coleman:1974jh,Linde:1976qh,Abbott:1975bn}, with Ref.~\cite{Abbott:1975bn,Linde:1976qh} pointing out that the tachyon (Landau's ghost) found in Ref.~
    \cite{Coleman:1974jh} simply was a consequence of expanding around the wrong vacuum, namely the $m=0$ solution in (\ref{twosolutions}). 
    \item
      There are mathematical proofs of triviality of scalar field theories in four dimensions, in particular by Aizenman and Duminil-Copin in Ref.~\cite{Aizenman:2019yuo}. Note that these proofs are limited to $N\leq 2$ and positive bare coupling, so they do not apply to the O(N) model in the large N limit. Using analytic continuation of the path integral contour, it is possible (but numerically challenging) to study negative coupling field theory on the lattice \cite{Romatschke:2023sce}.
    \item
      The proof by Coleman and Gross \cite{Coleman:1973sx} that only non-abelian gauge theories  in four dimensions can have asymptotic freedom rests on the same assumption as quantum triviality, namely that the bare coupling is positive.
  \item
    Scalar field theory with negative coupling was considered a long time ago by Symanzik \cite{Symanzik:1973hx}. For quantum mechanics, there is a whole literature surrounding negative coupling Hamiltonians which was opened up by Bender and B\"ottcher in Ref.~\cite{Bender:1998ke}. In quantum mechanics, strong numerical evidence for the equivalence of so-called PT-symmetric spectra and contour-deformed partition functions can be obtained \cite{Lawrence:2023woz}.
  \end{itemize}

  \subsection*{Homework Problems Lecture 3}
  \begin{enumerate}
  \item[3.1]
    Rederive the running coupling (\ref{runningL}) in cut-off regularization rather than dimensional regularization. You will need to also add a vacuum energy and a mass-counterterm.
  \item[3.2]
    Consider the 4d O(N) model with Euclidean action (\ref{4don}) at finite temperature. Calculate the finite-temperature corrections to the saddle point condition (\ref{saddleL3}) and show that real-valued solutions for $2 i \zeta_0=m^2$ of this equation cease to exist for
    \be
    \label{tc}
    T>T_c\simeq 0.616\Lambda_{\overline{\rm MS}}\,.
    \ee
  \item[3.3]
    Pretend that $\Lambda_{\overline{\rm MS}}$ in the O(N) model is the same as $\Lambda_{\overline{\rm MS}}$ in QCD. Use the particle data book to obtain values for $\Lambda_{\overline{\rm MS}}$ for QCD (for anything beyond one-loop, you will have to work backwards -- use the perturbative running coupling formula and $\alpha_s(M_Z)$ to find $\Lambda_{\overline{\rm MS}}$). Because the running $\alpha_s$ will need to touch the Z-mass $M_Z$, the number of ``light'' quarks should be $N_f=5$. Then, use the result (\ref{tc}) to estimate the deconfinement transition temperature of QCD. Compare to lattice QCD results for $T_c$ in QCD.
    \end{enumerate}

  \section*{Conclusions}

In these lectures, I have outlined the power of large N expansions for calculating physics observables in three different systems: quantum mechanics, non-relativistic fermions and relativistic scalar fields in four dimensions. These expansions rely on methods that have been developed a long time ago, but which remain underutilized in modern quantum field theory, especially in the context of in-medium problems such as finite temperature and finite density.

Large N methods are simply too powerful a tool to be relegated to the past: I sincerely hope that my lectures will be able to instill in other physicists the kind of excitement I felt when obtaining the neutron matter equation of state shown in Fig.~\ref{fig:three} or the Higgs mass prediction in Eq.~(\ref{dominant}), or the critical temperature for QCD in homework problem 3.3. 

Obviously, these lectures have barely scratched the surface of what can be done with large N techniques, with the ``guides to further reading'' in each chapter serving as little more than scattered sign posts in the vastness of the large N wonderland.

Therefore, stop reading, pick up your pencil, and start exploring the large N wonderland on your own! 
  
\section*{Acknowledgments}

I would like to thank the organizers of the 63rd Cracow School on Theoretical Physics, and especially Michał Praszałowicz, for their exceptional hospitality in Zakopane and for organizing such a lively and topical school. I also thank the students, postdocs, friends and colleagues at the Cracow School for many lively discussions and for engaging in my lectures, enduring my not-so-funny jokes with a kind smile, and actually working on my homework problems during this busy week of talks and lectures! I hope some of you will join me in the future in the Large N wonderland....

This work was supported by the Department of Energy, DOE award No DE-SC0017905. 

 \appendix

 \section{Numerically calculating the spectrum of quartic oscillator in multi-dimensional quantum mechanics}

In this appendix, I review a simple numerical scheme to solve for the eigenvalue spectrum (really mostly the ground-state energy $E_0$) of the Hamiltonian operator for quantum mechanics in $N$ dimensions,
\be
{\cal  H} =\frac{{\vec p}^{\,2}}{2}+\frac{\lambda}{N} \left(\vec{x}^{\, 2}\right)^2\,.
\ee
I assume a discrete eigenspectrum for the Hamiltonian ${\cal H}|n\rangle=E_n |n\rangle$. Using spherical coordinates, the angular part of the Laplace operator may be separated off whereas the radial part of the Schr\"odinger equation becomes
\be
\label{schroNd}
-\psi^{\prime\prime}(r)-\frac{N-1}{r}\psi^\prime(r)+\frac{l(l+N-2)}{r^2}\psi(r)+\frac{2 \lambda}{N} r^4 \psi(r)-2 E \psi(r)=0\,,
\ee
with $l$ the angular quantum number using the eigenvalues of the Laplacian on a $N-1$-dimensional sphere \cite[Eq.~(3.3)]{Grable:2022swa}. The boundary condition at $r=0$ for the wave function is
\be
\label{D1cond}
\lim_{r\rightarrow 0} r \psi(r)=0\,,
\ee
because otherwise $\vec{\nabla}^2 \left(\frac{1}{r}\right)=-4 \pi \delta(\vec{x})$ is not a solution to the Schr\"odinger equation. Rescaling of coordinates and energy values as
$r=\left(2\lambda\right)^{-\frac{1}{6}} \hat{r}$, $E=\left(2 \lambda\right)^{\frac{1}{3}}\hat E$, 
and rescaling $\psi$ as $\psi(\hat{r})=\frac{u(\hat{r})}{\sqrt{\hat r^{N-1}}}$, the Schr\"odinger equation becomes
\be
-u^{\prime\prime}(\hat r)+\frac{4l(l+N-2)+(N-1)(N-3)}{4\hat{r}^2}u(\hat r)+ \frac{\hat{r}^4}{N} u(\hat r)-2 \hat E u(r)=0\,.
\ee

For large $\hat{r}$, the $\hat{r}^4$ term in the potential dominates, so we choose a bounded wave-function by setting $u(\hat{r})=e^{-\frac{\hat{r}^3}{3 \sqrt{N}}}v(\hat{r})$, with $v(\hat{r})$ fulfilling
\be
-v^{\prime\prime}(\hat{r})+\frac{2 \hat{r}^2}{\sqrt{N}} v^\prime(\hat{r})+\left[\frac{4 l (l+N-2)+(N-1)(N-3)}{4 \hat{r}^2}+\frac{2 \hat{r}}{\sqrt{N}}-2\hat E\right]v(\hat{r})=0\,.
\ee

 \begin{table}
      \begin{tabular}{|c|cccccccccc|}
        \hline
        N & 1 & 2 &  3 & 4 & 5 & 6 & 7 & 8 & 9 & 10\\
        \hline
        $n=3$ & 0.209987 & 0.276277 & 0.500798 & 0.847447 & 1.30688 & 1.87549 & 2.55159 & 3.33429 & 4.22307 & 5.21761\\
        $n=7$ & 0.781176 & 0.591071 & 0.472416 & 0.429663 & 0.438973 & 0.486257 & 0.564257 & 0.669113 & 0.79867 &  0.951676\\
        $n=11$ & 0.657241 & 0.602466 & 0.583693 & 0.544103 & 0.498688 & 0.469752& 0.461883 & 0.47327 & 0.501472 & 0.544575\\
        $n=15$ & 0.668003 & 0.581889 & 0.552725 & 0.543857 & 0.536674 & 0.520139& 0.498165 & 0.480726 & 0.473021 & 0.476145\\
        $n=19$ & 0.668383 &  0.586494 & 0.551802 & 0.533517 & 0.524978 & 0.521291 & 0.516602 & 0.507002 & 0.494293 & 0.483397\\
        $n=23$ & 0.667887 & 0.586368 & 0.553458 & 0.534786 & 0.522767 & 0.51547&  0.511692 & 0.50933 &  0.505614 & 0.498988\\
        $n=27$ & 0.667991 & 0.586166 & 0.553281 & 0.535322 & 0.523688 & 0.5154 & 0.509576 & 0.505916 & 0.503775 & 0.501855\\
        $n=31$ & 0.66799 & 0.586204 & 0.553199 & 0.535197 & 0.523837 & 0.515885&  0.509889 & 0.505314 & 0.502048 & 0.499932\\        
        \hline
      \end{tabular}
      \caption{Estimates for the spectral gap $\frac{E_0}{\lambda^{\frac{1}{3}} N}$ for various values of $N$ resulting from solving $c_n=0$ for different approximation levels $n$. One should note that results stabilize as $n\rightarrow \infty$ as well as for $N\rightarrow \infty$.\label{tab:one}}
      \end{table}

The spectral gap is given by setting the angular quantum number to zero, $l=0$. It is then convenient to compactify the interval $\hat{r}\in [0,\infty)$ by introducing
  \be
  \hat{r}=\frac{y}{1-y}\,,\quad y\in[0,1)\,,\quad  v(\hat{r})=w(y)
    \ee
    and subsequently solving the Schr\"odinger equation by expanding $w(y)$ 
    in a power series in $y$. However, because of the boundary condition at $r=0$,  the series expansion must be taken as   
    \be
    \label{series}
    w(y)=y^{\frac{N-1}{2}}\sum_{n=0}^\infty c_n y^n\,.
    \ee
    The resulting recursion relation for the coefficients $c_n$ is somewhat unenlightening. For the first few coefficients we find
    \ba
    c_1&=&c_0\frac{ (N-1)}{2}\,,\nonumber\\
    c_2&=&c_0\frac{N^3-N-8 \hat{E}}{8 N}\,,\nonumber\\
    c_3&=&c_0\frac{N^4+3 N^3-N^2-3N+16 \sqrt{N}-24 \hat{E}(3+N)}{48 N}\,.
    \ea
    A simple yet effective way to obtain the spectrum $\hat{E}$ is by demanding that $c_n=0$ for sufficiently large $n$. For instance, setting $c_2=0$ leads to
    the crude estimate    $E_0^{(n=2)}=(2\lambda)^{\frac{1}{3}}\frac{N^3-N}{8}$ for the spectral gap. In practice, we find that the larger $N$, the higher $n$ needs to be in order for the spectral gap from $c_n=0$ to stabilize. Our results for the spectral gap for different $N,n$ are summarized in table \ref{tab:one}. One should note that the result for the spectral gap for the one component theory $N=1$ is consistent with the result from \cite[Eq.(IV.16)]{Hioe:1978jj}

    \section{Fixing Parameters in EFTs}
\label{sec:efts}
    
    For the case of $\slashed{\pi}$ EFT, the two-neutron parameter $C_0$ was identified with the s-wave scattering length in Eq.~(\ref{c0def}). In this appendix, I derive the corresponding relation for bosons. To this end, consider a simple example theory with an effective Lagrangian density
\begin{equation}
  \label{eq:effex0}
  {\cal L}=\phi\left(i\partial_t+\frac{\nabla^2}{2 M}\right)\phi-\frac{2 C_0}{4!}\phi^4\,,
\end{equation}
where for illustrative purposes we take $\phi$ to be a boson. The Lagrangian obeys Galilean invariance, and corresponds to an interacting non-relativistic field theory if $C_0$ is non-vanishing.

A standard calculation in quantum-field theory is the S-matrix
\begin{equation}
  S=1+i T\,,
\end{equation}
where the interaction part $T$ (also referred to as ``T-matrix'') may be expressed in terms of Feynman diagrams, see for example section 4.6 in Ref.~\cite{Peskin:1995ev}. Let us consider two-particle scattering: dividing ${\cal L}$ into a free field theory part ${\cal L}_0=\phi\left(i\partial_t+\frac{\nabla^2}{2 M}\right)\phi$ and an interaction part ${\cal L}_I={\cal L}-{\cal L}_0$, the T-matrix can be written as
\begin{equation}
T=\langle \phi_1 \phi_2 | e^{i \int d^4x {\cal L}_I} | \phi_A \phi_B\rangle_{\rm amputated,\, fully\, connected}\,,
\end{equation}
where time-ordering is implicit and the attributes ``amputated'' and ``fully connected'' refer to the class of Feynman diagrams contributing to $T$. Here $\phi_1$, $\phi_2$, $\phi_A$, $\phi_B$ are shorthand for the properties of the scattered particles, e.g. incoming particles 1 and 2, while A and B are outgoing particles. Examples for diagrams contributing to $T$ are
\begin{equation}
T=\vcenter{\includegraphics[width=0.8\linewidth]{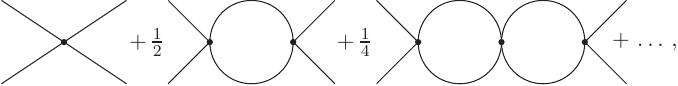}}
\end{equation}
where the symmetry factors for the diagrams have been made explicit and the Feynman rules in momentum space are:
\begin{itemize}
  \item
    There is a factor of $-2 i C_0$ for every vertex
    \item
  Energy and momentum are conserved at each vertex: \hbox{$(2 \pi)^4\delta(E_{\rm in}-E_{\rm out})\delta^3({\bf p}_{\rm in}-{\bf p}_{\rm out})$}
\item
  Integrate over every loop momentum: $\int \frac{d^4 p}{(2 \pi)^4}$
\item
  Each propagator is given by $\Delta(E,{\bf p})$ with $E,{\bf p}$ positive in the direction of momentum flow
\item
  All external lines are set to unity
  \end{itemize}
The propagator $\Delta(E,{\bf p})$ in momentum space may be calculated by performing a Fourier transform $\phi(x)=\int \frac{dE d{\bf p}}{(2 \pi)^4}e^{-i E t+i {\bf p}\cdot {\bf x}} \phi(E,{\bf p})$ in
\begin{eqnarray}
  &  - i\int d^4x {\cal L}_0=-i \int \frac{dE d{\bf p}}{(2 \pi)^4} |\phi(E,{\bf p})|^2 \left(E-\frac{{\bf p}^2}{2 M}\right)=\frac{dE d{\bf p}}{(2 \pi)^4} |\phi(E,{\bf p})|^2 \Delta^{-1}(E,{\bf p})&\,,\nonumber\\
  &  \Delta(E,{\bf p})=\frac{i}{E-\frac{{\bf p}^2}{2 M}+i 0^+}\equiv \Delta(P)\,,
  \end{eqnarray}
where we collectively denote four-momenta as $P\equiv(E,{\bf p})$.

With these set of Feynman rules, the T-matrix for the above diagrams can be evaluated. One finds that there is an overall factor of momentum conservation, which for the two-particle scattering case at hand implies
\begin{equation}
  i T=(2 \pi)^4\delta(E_1+E_2-E_A-E_B)\delta^3\left({\bf p}_1+{\bf p}_2-{\bf p}_A-{\bf p}_B\right) i {\cal M}\,.
\end{equation}
(Note that our normalization convention differs from standard relativistic quantum field theory, cf. Ref.~\cite{Peskin:1995ev}, but this difference does not play a role for the results found below). Here ${\cal M}$ is the scattering amplitude as defined in quantum field theory, and for the set of diagrams given in , 
it is given by
\begin{eqnarray}
  \label{eq:tmatrix}
  i {\cal M} &=& 2 (-i C_0)+2(-i C_0)^2\int \frac{d^4P}{(2\pi)^4}
    \Delta(P)\Delta(P_1+P_2-P)+\ldots\,.
\end{eqnarray}
It is convenient to evaluate ${\cal M}$ in the center of mass frame, e.g. $E_1=E_2=E_A=E_B=\frac{E}{2}$, ${\bf p}_1=-{\bf p}_2={\bf k}$, ${\bf p}_A=-{\bf p}_B={\bf k}^\prime$. Because these particles are on-shell, $E=\frac{{\bf k}^2}{M}=\frac{{\bf k}^{\prime 2}}{M}$. With these choices, the relevant loop integral in the scattering amplitude becomes
\begin{eqnarray}
  \label{eq:proi}
\int\frac{dp_0 d^3{\bf p}}{(2 \pi)^4}\Delta(p_0,{\bf p})\Delta(E-p_0,-{\bf p})=i \int \frac{d^{3}{\bf p}}{(2 \pi)^3}\frac{1}{E-\frac{{\bf p}^2}{M}+ i 0^+}\,.
  \end{eqnarray}
The integral is linearly divergent, so a regularization scheme has to be chosen. We will follow Ref.~\cite{Furnstahl:2008df} by employing dimensional regularization where $D=3\rightarrow D=3-2\epsilon$ such that 
\begin{eqnarray}
   \int \frac{d^D {\bf p}}{(2 \pi)^D}\frac{1}{{\bf p}^2-k^2- i 0^+}
  &=& \frac{1}{(4 \pi)^{D/2}}\Gamma(1-\frac{D}{2}) (-k^2)^{D/2-1}\,,\nonumber\\
  &=_{D\rightarrow 3}&-\frac{i k}{4 \pi}\,.
  \end{eqnarray}
Therefore, the scattering amplitude in the center of mass frame is given by
\begin{equation}
  {\cal M}=-2 C_0-2 C_0^2 M \frac{ik}{4\pi}+\ldots\,.
\end{equation}

Now let us redo the calculation in the context of the Schr\"odinger equation for two-particle scattering. For two particles with mass $M$ interacting with a two-body potential $V$, the Hamiltonian is given by
\begin{equation}
  {\cal H}=\frac{2  p^2}{2 M}+\hat V=\frac{ p^2}{M}+ V\,,
\end{equation}
where ${\cal H}_0=\frac{p^2}{M}$ is the free Hamiltonian. The free retarded Greens function operator is given by
\begin{equation}
G_0=\frac{1}{E-{\cal H}_0+i 0^+}\,,
\end{equation}
which can be used to write a solution to the full time-independent Schr\"odinger equation ${\cal  H} |\phi\rangle=E|\phi\rangle$ as
\begin{equation}
  \label{eq:schrosol}
  |\phi\rangle=|{\bf k}\rangle+ G_0 V |\phi\rangle\,,
\end{equation}
where $|{\bf k}\rangle$ is the solution to the free Schr\"odinger equation which we take to be normalized as $\langle x|{\bf k}\rangle=e^{i {\bf k}\cdot {\bf x}}$. The free retarded Greens function may be calculated with standard methods, finding
\begin{equation}
  \label{eq:gfform}
  \langle x | G_0 | x^\prime\rangle=-\frac{M}{4\pi}\frac{e^{i k |{\bf x}-{\bf x}^\prime|}}{|{\bf x}-{\bf x}^\prime|}\,,\quad
  \langle {\bf p}^\prime| G_0|{\bf p}\rangle=\frac{(2 \pi)^3 \delta^3({\bf p}-{\bf p}^\prime)}{E-p^2/M+i 0^+}\,,
  \end{equation}
such that the solution (\ref{eq:schrosol}) to the full Schr\"odinger equation for short range potentials $V$ becomes
\begin{equation}
  \phi(x)=e^{i {\bf k}\cdot {\bf x}}-\frac{M}{4\pi} \frac{e^{i k |{\bf x}|}}{|{\bf x}|}\int d^3x^\prime e^{-i {\bf k}^\prime \cdot {\bf x}^\prime} V(x^\prime) \phi(x^\prime)\,,
\end{equation}
where ${\bf k}^\prime\equiv k \frac{\bf x}{|{\bf x}|}$. This form may be compared to that of a scattered wave with scattering amplitude $f({\bf k},{\bf k}^\prime)$:
\begin{equation}
  \phi(x)=e^{i {\bf k}\cdot {\bf x}}+\frac{e^{i k |{\bf x}|}}{|{\bf x}|}f({\bf k},{\bf k}^\prime)\,,
  \end{equation}
from which it follows that
\begin{equation}
  \label{eq:schroam}
  f({\bf k},{\bf k}^\prime)=-\frac{M}{4\pi}\langle {\bf k}^\prime | V | \phi\rangle\,.
\end{equation}
We will find that the scattering amplitude $f$ as used in the Schr\"odinger equation is related to the scattering amplitude ${\cal M}$ calculated in quantum field theory (\ref{eq:tmatrix}) up to a normalization. For a spherically symmetric scattering potential, the scattering amplitude may be decomposed entirely in partial waves as
\begin{equation}
  \label{eq:partialwaves}
  f({\bf k},{\bf k}^\prime)=\sum_{l=0}^\infty \frac{(2 l+1)P_l(\cos\theta)}{k \cot \delta_l(k)-i k}\,,
\end{equation}
where ${\bf k}\cdot {\bf k}^\prime=k^2 \cos\theta$ and $\delta_l(k)$ are the energy-dependent scattering phase shifts. For low energy scattering $k\rightarrow 0$, the higher partial waves are suppressed and s-wave scattering $l=0$ dominates the scattering amplitude. One finds that in this case, the form of the s-wave phase shift is universally given by
\begin{equation}
  k \cot \delta_0(k)=-\frac{1}{a_0}+\frac{r_0}{2}k^2+{\cal O}(k^3)\,,
\end{equation}
where $a_0,r_0$ are the s-wave scattering length and effective range, respectively. The scattering length and effective range are reliably measured experimentally for a variety of systems.

Using again the result for the scattering amplitude in the Schr\"odinger calculation given in Eq.~(\ref{eq:schroam}), where $|\phi\rangle$ is given by Eq.~(\ref{eq:schrosol}), we have
\begin{equation}
  f=-\frac{M}{4\pi}\left(\langle {\bf k}^\prime| V|{\bf k}\rangle+\langle {\bf k}^\prime|V  G_0  V|{\bf k}\rangle+\ldots\right)\,.
  \end{equation}
Using $\langle {\bf k}^\prime|V|{\bf k}\rangle=V({\bf k},{\bf k}^\prime)$ and the known form of the Green's function (\ref{eq:gfform}) leads to 
\begin{eqnarray}
  \label{eq:schrofin}
  f&=&-\frac{M}{4\pi}\left(V({\bf k}^\prime,{\bf k})+\int \frac{d^3 p}{(2 \pi)^3}
  V({\bf k}^\prime,{\bf p})\frac{1}{E-p^2/M+i 0^+}V({\bf p},{\bf k})+\ldots\right)\,.
  \end{eqnarray}
Comparing (\ref{eq:schrofin}) to Eq.~(\ref{eq:tmatrix}) when using (\ref{eq:proi}), one finds that the structure of the integrals is very similar. In fact one finds that
\begin{equation}
  f=\frac{M}{4\pi}\frac{\cal M}{2}
  \end{equation}
if $V({\bf p},{\bf q})=C_0$ such that if we focus on low-energy (s-wave) scattering, we have
\begin{equation}
  \label{eq:matching1}
  \frac{4\pi}{M}\frac{1}{-\frac{1}{a_0}-ik +\frac{r_0}{2}k^2+\ldots}=-C_0+C_0^2 M \frac{i k}{4\pi}+\ldots \,,
\end{equation}
which implies
\begin{equation}
  \label{eq:mach2}
  C_0=\frac{4 \pi a_0}{M} \,.
\end{equation}
Note that the matching includes the term linear in $k$ in (\ref{eq:matching1}) which is a non-trivial consistency check. Equation (\ref{eq:mach2}) implies that we have matched the leading low-energy constant $C_0$ to an experimentally measurable quantity, the scattering length $a_0$. 

\newpage

\section*{Solution to Homework Problems}
\subsection*{Solution Problem 1.1}

One way to calculate the perturbative ground state energy $E_0$ is to write down the perturbative expansion for the path integral for the partition function
\be
Z=\int {\cal D}\phi e^{-\int_0^\beta d\tau \frac{\dot \phi^2}{2}}
\left(1-\lambda \int_0^\beta d\tau \phi^4(\tau)+\ldots\right)
\ee
and use Wick's theorem to rewrite this as
\be
Z=Z_0\left(1-\lambda \int_0^\beta 3 G^2(0)+\ldots\right)\,,
\ee
where
\be
Z_0=\int {\cal D}\phi e^{-\int_0^\beta d\tau \frac{\dot \phi^2}{2}}\,,\quad
G(\tau)=\langle \phi(\tau)\phi(0)\rangle_0\,.
\ee
We can read off the propagator from the 'free' action $S_0=\int_0^\beta d\tau \frac{\dot \phi^2}{2}$ as
\be
G(\tau)=T\sum_{n} \frac{e^{i \omega_n \tau}}{\omega_n^2}\,,
\ee
with $\omega_n=2 \pi n T$ and $n\in \mathbb{Z}$ the bosonic Matsubara frequency.
Because of the zero mode $\omega_0=0$, this expression is ill-defined, so our naive result is $G(0)=\infty$, corresponding to a diverging result for $E_0$.

However, we could try to do better by explicitly ignoring the zero mode, so that
\be
G(0)=2 T \sum_{n=1}^\infty \frac{1}{\omega_n^2}=\frac{2\beta}{(2\pi)^2}\zeta(2)=\frac{\beta}{12}\,,
\ee
so that
\be
Z=Z_0\left(1-\frac{3 \lambda \beta^3}{144}+\ldots\right)\,.
\ee
Recasting this as
\be
Z=Z_0 e^{-\frac{3 \lambda \beta^3}{144}}\,,
\ee
and comparing to the expected result $Z\propto e^{-\beta E_0}$ in the low temperature limit, we get $E_0=\lim_{\beta\rightarrow \infty}\frac{3 \lambda \beta^2}{144}\rightarrow \infty$, which is still not a sensible result.

The upshot is that there is no simple way to calculate $E_0$ in perturbation theory for the pure quartic oscillator. This problem does not have a perturbative solution! This could have been anticipated from the known scaling behavior $E_0\propto \lambda^{\frac{1}{3}}$, which does not easily lend itself to a nice power series in $\lambda$. To obtain even the simplest finite result for $E_0$ requires resumming an infinite number of perturbative diagrams. 

\subsection*{Solution Problem 1.2}

Starting with the auxiliary field action
\be
S_E=\frac{N}{2}{\rm Tr}\ln \left[-\partial_\tau^2+2 i \zeta\right]+\int_0^\beta \frac{N \zeta^2}{4\lambda}\,,
\ee
we first perform a split into a constant mode $\zeta_0$ and fluctuations $\zeta^\prime$:
\be
\zeta(\tau)=\zeta_0+\zeta^\prime(\tau)\,,
\ee
with $\int_0^\beta d\tau \zeta^\prime(\tau)=0$. Next it is useful to convert the logarithm of the differential operator to frequency space. This is easily done by recognizing that the whole expression arose from integrating out the scalar fields
\be
S_\phi=\int_0^\beta d\tau \vec{\phi}(\tau)\left[-\partial_\tau^2+2 i \zeta_0+2 i \zeta^\prime(\tau)\right]\vec{\phi}(\tau)\,.
\ee
Fourier-transforming the fields
\be
\vec{\phi}(\tau)=T \sum_n e^{i \omega_n \tau}\tilde \phi_n\,,
\ee
with the bosonic Matsubara frequencies $\omega_n=2 \pi n T$ and $n\in \mathbb{Z}$ leads to
\be
S_\phi=T \sum_{i j} \tilde \phi_i \left[\delta_{i,-j}\left(\omega_i^2+2 i \zeta_0\right)+2 T i \zeta_{i+j}\right]\tilde \phi_j\,. 
\ee
Letting $j\rightarrow -j$, we thus can write the action in matrix form
\be
S_\phi=T \sum_{i j} \tilde \phi_i \left[\delta_{ij}\left(\omega_i^2+2 i \zeta_0\right)+2 T i \zeta^\prime_{i-j}\right]\tilde \phi_j\,,
\ee
and hence the auxiliary field action in Fourier space becomes
\be
S_E=\frac{N}{2}{\rm Tr}\ln\left[\delta_{ij}\left(\omega_i^2+2 i \zeta_0\right)+2 T i \zeta^\prime_{i-j}\right]+\frac{N \beta \zeta_0^2}{4\lambda}+\frac{N T}{4\lambda}\sum_i \zeta^\prime_i \zeta^\prime_{-i}\,.
\ee
Expanding the logarithm generates terms of order $\zeta^{\prime 0},\zeta^{\prime 2},\zeta^{\prime 3},\zeta^{\prime 4},\ldots$,
\ba
\label{exp}
   {\rm Tr}\ln\left[\omega_i^2+2 i \zeta_0\right]+{\rm Tr}\ln\left[\delta_{ij}+\frac{2 T i}{\omega_i^2+2 i \zeta_0} \zeta^\prime_{i-j}\right]&=&\sum_i \ln \left[\omega_i^2+2 i \zeta_0\right]\\
   -\frac{1}{2}(2 T i)^2 \sum_{i,j}G_i \zeta^\prime_{i-j}G_j \zeta^\prime_{j-i}
   &+&\frac{1}{3}(2 T i)^3 \sum_{ijk}G_i \zeta^\prime_{i-j}G_j \zeta^\prime_{j-k}G_k \zeta^\prime_{k-i}+\ldots\nonumber
   \ea
   where $G_i\equiv \frac{1}{\omega_i^2+2 i \zeta_0}$, and the term linear in $\zeta^\prime$ vanishes because $\zeta^\prime_0=0$ as a consequence of $\int d\tau \zeta^\prime(\tau)=0$. Shifting the index $i\rightarrow i+j$, the quadratic term becomes in the zero temperature limit
   \be
   2 T \sum_{i} \zeta^\prime_i \zeta^\prime_{-i} T \sum_j G_j G_{i+j} \rightarrow
   2 \int \frac{dk}{2\pi} |\zeta^\prime(k)|^2\int \frac{dp}{2\pi} \frac{1}{p^2+2 i \zeta_0}\frac{1}{(p+k)^2+2 i \zeta_0}\,,
   \ee
   which is the expression quoted for the Gaussian fluctuations in lecture 1.

   The cubic, quartic and higher order terms in $\zeta^\prime$ in the expansion (\ref{exp}) can be treated by expanding the exponential, symbolically as
   \be
   e^{-x^3-x^4-x^5+\ldots}=1-x^3-x^4-x^5+\ldots\,,
   \ee
   and noting that all odd terms in this expansion vanish because the auxiliary field action is quadratic in $\zeta^\prime$. Hence the first non-vanishing correction to the Gaussian fluctuations comes from the quartic term in (\ref{exp}), given by
   \be
   -\frac{1}{4}(2 T i)^4 \sum_{ijkl} G_i G_j G_k G_l \zeta^\prime_{i-j}\zeta^\prime_{j-k}\zeta^\prime_{k-l}\zeta^\prime_{l-i}\,,
   \ee
   which together with the leading factor $\frac{N}{2}$ contributes a factor of 
   \be
   \label{co3}
   C_3=1+2 N T^4 \sum_{ijkl} G_i G_j G_k G_l \langle \zeta^\prime_{i-j}\zeta^\prime_{j-k}\zeta^\prime_{k-l}\zeta^\prime_{l-i}\rangle\,,
   \ee
   to the partition function $Z$, where $\langle \rangle$ correspond to expectation values wrt to the Gaussian action for the fluctuations. Since the propagator for the auxiliary fields from this action is
   \be
   \langle \zeta^\prime_i \zeta^\prime_j\rangle =\frac{\beta\delta_{i,-j}}{N}D_i\,, \quad D_i=\frac{1}{\frac{1}{2\lambda}+4 \Pi_i}\,, \quad \Pi_i=\frac{T}{2}\sum_j G_j G_{i+j}\,.
   \ee
   Using this in (\ref{co3}) leads to
   \be
   C_3=1+\frac{2 T^2}{N} \sum_{ijkl} G_i G_j G_k G_l \left(D_{i-j} D_{k-l}\delta_{i,k}+D_{i-j} D_{j-k}\delta_{i,j+l-k}+D_{i-j}D_{j-k}\delta_{l,j}\right)\,,
   \ee
   or after rearranging
   \be
   C_3=1+\frac{2 \beta}{N}\left(2 T \sum_i G_i^2 S_i^2 +T^2 \sum_{ik} G_k G_i D_{i-k} V_{i,k}\right)\,,
   \ee
   where
   \be
   S_i=T \sum_j G_j D_{i-j}\,, \quad V_{i,k}=T \sum_j G_{j+i-k} G_j D_{j-k}\,.
   \ee
   In the zero temperature limit, we have
   \be
   \label{c3}
    C_3=1+\frac{2 \beta}{N}\left(2 \int \frac{dk}{2\pi} \left(G(k) S(k)\right)^2 +\int \frac{dk dp}{(2\pi)^2} G(k) G(p) D(p-k) V(p,k)\right)\,,
    \ee
    with
    \be
    S(p)=\int\frac{dk}{2\pi} G(k) D(p-k)\,,\quad
    V(p,k)=\int \frac{dq}{2\pi}G(q+p-k)G(q)D(q-k)\,,
    \ee
    and
    \be
    G(k)=\frac{1}{k^2+m^2}\,,\quad D(k)=\frac{1}{\frac{1}{2\lambda}+4\Pi(k)}\,, \quad m^2=2 i \zeta_0\,.
    \ee

    The calculation of $\Pi(k)$ was done in lecture 1 as (\ref{piqm}, giving
    \be
    \Pi(k)=\frac{1}{2 m \left(k^2+4 m^2\right)}\,,
    \ee
    so that the partition function to NNLO large N becomes
    \be
    Z=\int d\zeta_0 e^{-\frac{N \beta m}{2}+\frac{N \beta m^4}{16\lambda}-\frac{\beta}{2}\int \frac{dk}{2\pi} \ln\left(1+8\lambda \Pi(k)\right)}C_3\,,\quad m^2=2 i \zeta_0\,. 
    \ee
     The integral can be evaluated as
    \ba
    \int \frac{dk}{2\pi}\ln\left(1+8 \lambda \Pi(k)\right)&=&m \int \frac{dk}{2\pi}\left[\ln\left(k^2+4+\frac{4\lambda}{m^3}\right)-\ln\left(k^2+4\right)\right]\,,\nonumber\\
    &=&m \left[\sqrt{4+ \frac{4\lambda}{m^3}}-2\right]\,.
\ea
    
    For $C_3$, we need
    \ba
    S(p m)&=&\frac{2\lambda}{m}\int \frac{dk}{2\pi}\frac{k^2+4}{\left((k-p)^2+1\right)\left(k^2+4+\frac{4\lambda}{m^3}\right)}\,,\nonumber\\
    &=&\frac{2\lambda}{m} \left[\frac{1}{2}-\frac{2 \lambda}{m^3}\frac{2+\frac{1}{\sqrt{1+\frac{\lambda}{m^3}}}}{p^2+4\left(1+\frac{\lambda}{m^3}\right)+4 \sqrt{1+\frac{\lambda}{m^3}}+1}\right]\,.
    \ea
    The remaining integral could be done analytically, but since I'm lazy, I will evaluate it numerically for the LO mass parameter $m=(2\lambda)^{\frac{1}{3}}$. We get
    \ba
    \label{refvals}
    \int \frac{dk}{2\pi}\left(G(k)S(k)\right)^2&\simeq & m \times 0.0375945\ldots\,,\\
    \int \frac{dk dp}{(2\pi)^2} G(k) G(p) D(p-k) V(p,k)&\simeq & m \times 0.0168984\ldots\,,\nonumber
    \ea
    
At large N, the integral over $\zeta_0$ is dominated by the saddle, which we need to NNLO in large N. The saddle point condition becomes
    \be
    \frac{1}{2}-\frac{m^3}{4\lambda}+\frac{1}{2 N} \frac{d}{dm} \left[m\sqrt{4+ \frac{4\lambda}{m^3}}-2 m\right]=0\,,
    \ee
    or
    \be
    m=(2\lambda)^{\frac{1}{3}}\left(1+\frac{1}{N} \left[m\sqrt{4+ \frac{4\lambda}{m^3}}-2 m\right]\right)^{\frac{1}{3}}
    \ee

    Expanding in powers of $\frac{1}{N}$, we get for the mass parameter to NLO
    \be
    m=(2\lambda)^{\frac{1}{3}}\left(1+\frac{-4+\sqrt{6}}{6N}\right)\,,
    \ee
    so that
    \be
    -\frac{\ln Z}{\beta}=(2\lambda)^{\frac{1}{3}}\times\left(\frac{3 N}{8}+\frac{\sqrt{6}-2}{2}+\frac{11-4 \sqrt{6}}{24 N}-\frac{0.18418}{N}\right)\,,
    \ee
    where I have used the numerical values (\ref{refvals}) for $C_3$ given by (\ref{c3}).

    Putting everything together, we have
    \be
    E_0\simeq (2\lambda)^\frac{1}{3}\left(\frac{3 N}{8}+\frac{\sqrt{6}-2}{2}-\frac{0.134}{N}\right)\,,
    \ee
    so that the NNLO contribution to the ground state energy is
    \be
    E_0^{NNLO}\simeq- 0.1689\times N^{-1}\lambda^{\frac{1}{3}}\,.
    \ee

\subsection*{Solution Problem 1.3}

Following the same steps as in the lecture for QM, we obtain the expression for the path integral partition function
\be
Z=\int {\cal D}\zeta e^{-S_E}\,,\quad S_E=\frac{N}{2}{\rm Tr}\ln \left[-\partial_\mu\partial_\mu+2 i \zeta\right]+\int_0^\beta d\tau \int d^2x \frac{N \zeta^2}{4\lambda}\,.
\ee
Splitting the auxiliary field into a constant $\zeta_0$ and fluctuations, the LO large N expression once again does not include the contribution from the fluctuations, hence
\be
\lim_{N\gg 1}Z=\int d\zeta_0 e^{-S_{R0}}\,,\quad S_{R0}=\frac{N}{2}{\rm Tr}\ln \left[-\partial_\mu\partial_\mu+2 i \zeta_0\right]+\frac{N \beta V \zeta_0^2}{4\lambda}\,,
\ee
where $V=\int d^2x$ is the ``volume'' (area) of space. We now need to calculate the trace of the logarithm, which can be done using the thermal sum result in lecture 1 as follows:
\ba
   {\rm Tr}\ln\left[-\partial_\mu \partial_\mu+m^2\right]&=&V \sum_n \int \frac{d^2k}{(2\pi)^2}\ln\left[\omega_n^2+{\bf k}^2+m^2\right]\,,\nonumber\\
   &=&2 V \int \frac{d^2k}{(2\pi)^2}\ln\left[2 \sinh\left(\frac{\sqrt{{\bf k}^2+m^2}}{2T}\right)\right]\,,\\
   &=&\beta V \int \frac{d^2k}{(2\pi)^2}\sqrt{{\bf k}^2+m^2}+2 V \int \frac{d^2k}{(2\pi)^2}\ln\left[1-e^{-\frac{\sqrt{{\bf k}^2+m^2}}{T}}\right]\,.\nonumber
   \ea
   Now using dimensional regularization to evaluate
   \be
   \int \frac{d^2k}{(2\pi)^2}\sqrt{{\bf k}^2+m^2}=\frac{1}{(4\pi)}\frac{\Gamma\left(-\frac{3}{2}\right)}{\Gamma\left(-\frac{1}{2}\right)} m^3=-\frac{m^3}{6\pi}\,.
   \ee
   In terms of $m^2=2 i \zeta_0$, the R0-action therefore is
   \be
   \frac{S_{R0}}{N \beta V}=-\frac{ m^3}{12 \pi} +T \int \frac{d^2k}{(2\pi)^2}\ln\left[1-e^{-\frac{\sqrt{{\bf k}^2+m^2}}{T}}\right]-\frac{m^4}{16\lambda}\,,
   \ee
   and the saddle point condition becomes
   \be
   0=-\frac{ m^2}{4 \pi} +m \int \frac{d^2k}{(2\pi)^2}\frac{n_B(\sqrt{k^2+m^2})}{\sqrt{k^2+m^2}}-\frac{m^3}{4\lambda}\,,
   \ee
   where $n_B(k)=\frac{1}{e^{\beta k}-1}$ is the Bose-Einstein thermal distribution factor. The remaining integral can be evaluated as
   \be
   \int \frac{d^2k}{(2\pi)^2}\frac{n_B(\sqrt{k^2+m^2})}{\sqrt{k^2+m^2}}=\sum_{n=1}^\infty \int \frac{dk}{2\pi} \frac{ke^{-n \beta \sqrt{k^2+m^2}} }{\sqrt{k^2+m^2}}=-\frac{T}{2\pi} \ln \left(1-e^{-\beta m}\right)\,.
   \ee
   We distinguish the cases of the free theory $\lambda=0$, where $m=0$, and the strong coupling limit $\lambda\rightarrow \infty$, for which $m$ has to fulfill
   \be
    \frac{m\beta}{4 \pi} +\frac{1}{2\pi} \ln \left(1-e^{-\beta m}\right)=0\,.
    \ee
    This equation has the curious solution
    \be
    \label{curious}
    m\beta=2 \ln \frac{1+\sqrt{5}}{2}\,.
    \ee

    The pressure $p$ and entropy density $s$ for the theory are given as
    \be
    p=-\frac{S_{R0}}{\beta V}\,,\quad s=\frac{d p}{dT}\,,
    \ee
    where $p$ is evaluated at the stationary point (solution for the saddle point condition). Since $p$ contains both explicit T-dependencies as well as implicit dependencies (e.g those through the T-dependence of $m$), we split
    \be
    s=\frac{\partial p}{\partial T}+\frac{\partial p}{\partial m} \frac{dm}{dT}\,.
    \ee
    However, since $\frac{\partial p}{\partial m}\propto \frac{\partial S_{R0}}{\partial m}=0$ for the saddle point, we simply have
    \ba
    s&=&\frac{\partial p}{\partial T}=N \int \frac{d^2k}{(2\pi)^2}\ln\left[1-e^{-\frac{\sqrt{{\bf k}^2+m^2}}{T}}\right]-N \beta \int \frac{d^2k}{(2\pi)^2}\sqrt{k^2+m^2} n_B(\sqrt{k^2+m^2})\,,\nonumber\\
    &=&\frac{N\beta}{4\pi} \int_m^\infty dx\,  (3x^2-m^2) n_B(x)\,,
    \ea
    where I've used integration by parts to obtain the last result.

    For the free theory case, $m=0$, and one finds
    \be
    s_{\rm free}=\frac{3 N\beta}{4\pi} \sum_{n=1}^\infty \frac{2}{n^3 \beta^3}=\frac{3 N T^2 \zeta(3)}{2 \pi}\,.
    \ee
    By contrast, for the strongly coupled theory one has
    \ba
    s_{\infty}&=&\frac{N\beta}{4\pi} \int_m^\infty dx\,  (3x^2-m^2) n_B(x)\,,\\
    &=&\frac{N T^2}{2\pi}\left[3 {\rm Li}_3\left(e^{-\beta m}\right)+3 \beta m {\rm Li}_2\left(e^{-\beta m}\right)-(\beta m)^2\ln\left(1-e^{-\beta m}\right)\right]\,.
    \ea
    Noting that for the saddle point (\ref{curious}) one can use the identities
    \ba
    \ln\left(1-e^{-\beta m}\right)&=&-\frac{\beta m}{2}\,,\nonumber\\
            {\rm Li}_2\left(e^{-\beta m}\right)&=&\frac{\pi^2}{15}-\frac{(\beta m)^2}{4}\,,\nonumber\\
            {\rm Li}_3\left(e^{-\beta m}\right)&=&\frac{4}{5}\zeta(3)-\frac{\pi^2 \beta m}{15}+\frac{(\beta m)^3}{12}\,,
            \ea
            and hence one finds
            \be
            s_{\infty}=\frac{12 N T^2 \zeta(3)}{10\pi}=\frac{4}{5}s_{\rm free}\,,
            \ee
            matching the finding in Ref.~\cite{Romatschke:2019ybu}.

\subsection*{Solution Problem 1.4}

We start with the auxiliary-field formulation of the partition function also given in lecture 1:
  \be
  Z=\int {\cal D}\vec{\phi}{\cal D}\zeta e^{-\int d^3x\left[ \frac{1}{2}\vec{\phi}\left(-\Box+2 i \zeta\right)\vec{\phi} +\frac{N}{4\lambda}\zeta^2\right] }\,.
  \ee
  For further convenience, we rescale $\zeta\rightarrow \frac{\zeta}{2}$ to find
  \be
  Z=\int {\cal D}\vec{\phi}{\cal D}\zeta e^{-\int d^3x\left[ \frac{1}{2}\vec{\phi}\left(-\Box+ i \zeta\right)\vec{\phi} +\frac{N}{16\lambda}\zeta^2\right] }\,.
  \ee

  We can always rewrite the field $\zeta=\zeta_0+\zeta^\prime$, where $\zeta_0$ is the global zero mode and $\zeta^\prime$ are fluctuations around that zero mode. Ignoring the contributions from the fluctuations gives the R0 partition function
  \be
  Z_{\rm R0}=\int {\cal D}\vec{\phi}\int d \zeta_0 e^{-\int d^3x\left[ \frac{1}{2}\vec{\phi}\left(-\Box+ i \zeta_0\right)\vec{\phi} +\frac{N}{16\lambda}\zeta_0^2\right] }\,,
  \ee
  where the integral over $\zeta_0$ is a \textit{normal} integral (not a path integral!). For R0, the integral over the fields $\vec{\phi}$ is quadratic, so the integral can be done in closed form. One finds
  \be
  Z_{\rm R0}=\int d\zeta_0 e^{\beta V N P(\sqrt{i \zeta_0})}\,,
  \ee
  where at zero temperature
  \be
  P(m)=-\frac{1}{2}\int \frac{d^3k}{(2\pi)^3} \ln(k^2+m^2)+\frac{m^4}{16\lambda}\,.
  \ee
  At large N, $Z_{R0}$ is given exactly by the method of steepest descent, e.g. the integral over $\zeta_0$ is evaluated from the location of the saddle. The saddle point condition is
  \be
  P^\prime(m)=0=-m\int \frac{d^3k}{(2\pi)^3} \frac{}{k^2+m^2}+\frac{m^3}{4\lambda}\,,
  \ee
  which is in the form of a gap equation

  In the strong coupling limit, the gap equation is
  \be
  -m\int \frac{d^3k}{(2\pi)^3}\frac{1}{k^2+m^2}=0\,.
  \ee
  The integral can be solved in dimensional regularization to give
  \be
  -m\left(-\frac{m}{4\pi}\right)=0\,,
  \ee
  which only has the solution $m=0$

  The full propagator in R0 is given by
  \be
  \langle \phi_i(x)\phi_j(0)\rangle=Z^{-1}\int d\zeta_0\int {\cal D}\vec{\phi} e^{-\int d^3x\left[ \frac{1}{2}\vec{\phi}\left(-\Box+ i \zeta_0\right)\vec{\phi} +\frac{N}{16\lambda}\zeta_0^2\right]}\phi_i(x)\phi_j(0)\,.
  \ee
  In Fourier space in the strong coupling limit then
  \be
  G_{R0}(k)=Z^{-1}\int d\zeta_0 e^{\beta V N P(\sqrt{i\zeta_0})}\frac{1}{k^2+i \zeta_0}\,.
  \ee
  In the large N limit, the integral gets evaluated through the method of steepest descent, fixing the value of $i\zeta_0$ at the solution of the gap equation. That value was calculated in the preceding step, and we find
  \be
  G_{R0}(k)=\frac{1}{k^2}\,.
  \ee
  From this, we find
  \be
  \eta_{\rm LO}=0\,.
  \ee

  To go beyond this result, we need to increase the large N accuracy by keeping sub-leading terms. To this end add and subtract terms
  \be
  \frac{1}{2}\int d^3k \vec{\phi}(-k) \Sigma(k) \vec{\phi}(k)\,,
  \ee
  and
  \be
  \frac{1}{2}\int d^3k \zeta^\prime(-k) \Pi(k) \zeta^\prime(k)\,,
  \ee
  to the R0 action. Since the same terms were added and subtracted, the QFT is unchanged. The functions $\Sigma,\Pi$ can be fixed by calculating $\langle \vec{\phi}(x)\vec{\phi}(0)\rangle$ and $\langle \zeta^\prime(x)\zeta^\prime(0)\rangle$ to leading non-trivial order in large N. You are welcome to do this calculation, but the end result (in x-space) should be 
  \be
  \label{R2}
  \Pi(x)=\frac{1}{2} G^2(x)\,,\quad \Sigma(x)= 4 D(x)G(x)\,,
  \ee
  where in Fourier space $D(k)=\frac{1}{N}\frac{1}{\frac{1}{2\lambda}+ 4\Pi(k)}$, 
  cf. Ref.~\cite{Romatschke:2019rjk} (note this reference uses different convention/notations such as $\Pi\leftrightarrow \Sigma$). The resulting action is called \textit{resummation level 2} or R2 for short.

  In R2, in Fourier space at zero temperature we have
  \be
  \Pi(k)=\frac{1}{2} \int \frac{d^3p}{(2\pi)^3}\frac{1}{p^2({\bf p}-{\bf k})^2}\,.
  \ee
  We can use Feynman parameters to rewrite this integral as
   \be
  \Pi(k)=\frac{1}{2} \int \frac{d^3p}{(2\pi)^3}\int_0^1 dx\frac{1}{\left[p^2 x+ (1-x)({\bf p}-{\bf k})^2\right]^2}\,.
  \ee
  Shifting the integration variable ${\bf p}$ and performing the integration over $p$ using dimensional regularization, we can write this as
  \be
  \Pi(k)=\frac{1}{2} \int \frac{d^3p}{(2\pi)^3}\int_0^1 dx\frac{1}{\left[p^2 + (1-x)x k^2\right]^2}=\frac{1}{16\pi}\int_0^1 dx \frac{1}{\sqrt{x(1-x)k^2}}=\frac{1}{16 \sqrt{k^2}} \,.
  \ee

  Using this result in the strong coupling expression (\ref{R2}) we get
  \be
  \Sigma(k)=\frac{1}{N} \int \frac{d^3p}{(2\pi)^3} \frac{G({\bf p}-{\bf k})}{\Pi(p)}=\frac{16}{N} \int \frac{d^3p}{(2\pi)^3} \frac{\sqrt{p^2}}{({\bf p}-{\bf k})^2}\,.
  \ee
  Feynman parameters are again used for this expression, giving after a similar shift of integration momenta as before
   \be
   \Sigma(k)=\frac{16}{N} \frac{\Gamma\left(\frac{1}{2}\right)}{\Gamma\left(-\frac{1}{2}\right)} \int \frac{d^3p}{(2\pi)^3}\int_0^1dx \frac{x^{-\frac{3}{2}}}{\left[p^2+x(1-x)k^2\right]^{1/2}}\,.
   \ee
   The remaining integral can again be calculated easily in dim-reg by replacing
   \be
   3\rightarrow 3-2\varepsilon\,.
   \ee
One finds
   \be
   \Sigma(k)=\frac{16}{N} \frac{\Gamma\left(\frac{1}{2}\right)}{\Gamma\left(-\frac{1}{2}\right)} \frac{\mu^{2\varepsilon}}{(4\pi)^{\frac{3}{2}-\varepsilon}} \frac{\Gamma(-1+\varepsilon)}{\Gamma\left(\frac{1}{2}\right)} \int_0^1dx x^{-\frac{3}{2}}\left[x(1-x)k^2\right]^{1-\varepsilon}\,.
   \ee
   Expanding this expression to ${\cal O}(\varepsilon^0)$ in small $\varepsilon$, one obtains
   \be
   \Sigma(k)=\frac{k^2}{N\pi^2} \int_0^1dx \sqrt{\frac{(1-x)^2}{x}} \left(\frac{1}{\varepsilon}-\ln (x(1-x))+\ln \frac{\bar\mu^2 e^{1}}{k^2}\right)  \,.
   \ee
   The remaining integral over x can be evaluated analytically, finding
   \be
   \Sigma(k)=\frac{4 k^2}{3 N\pi^2} \left(\frac{1}{\varepsilon}+\ln \frac{\bar\mu^2}{k^2}+\frac{16-6 \ln 2}{3}\right)  \,.
   \ee

   The formal expansion
   \be
   G^{-1}(k)=\left(k^2\right)^{1-\frac{\eta_{\rm LO}}{2}-\frac{\eta_{\rm NLO}}{2N}+\ldots}=(k^2)^{1-\frac{\eta_{\rm LO}}{2}}\left(1-\frac{\eta_{\rm NLO} k^2}{2N}\ln k^2+\ldots\right)\,,
   \ee
   should be compared to 
   \be
   G_{R2}^{-1}(k)=k^2+\Sigma(k)=k^2+\frac{4 k^2}{3 N\pi^2} \left(\frac{1}{\varepsilon}+\ln \frac{\bar\mu^2}{k^2}+\frac{16-6 \ln 2}{3}\right)+\ldots
   \ee
   Comparing the coefficient of the logarithmic term we find
   \be
   \eta_{\rm NLO}=\frac{8}{3 \pi^2}\,,
   \ee
    so that the critical exponent becomes
   \be
   \eta=\frac{8}{3 N \pi^2}+{\cal O}(\frac{1}{N^2})\,.
   \ee
   This matches the published result in Ref.~\cite{ma1973critical}

   \subsection*{Solution Problem 2.1}

   In the strong coupling limit, the pressure from (\ref{p3}) becomes
   \be
   \lim_{a_0\rightarrow -\infty}p(0,\Delta)\rightarrow \frac{2\mu }{5} \frac{(2 M \mu)^{\frac{3}{2}}}{3\pi^2} g\left(\frac{\mu}{\sqrt{\mu^2+\Delta^2}}\right)\,. 
   \ee
   Finding the minimum of this function wrt $\Delta$ is akin to finding the minimum of the function $g$, defined below (\ref{p3}) in terms of elliptic integrals:
   \be
g(y)=y^{-\frac{5}{2}}\left[(4 y^2-3)E\left(\frac{1+y}{2}\right)+\frac{3+y-4 y^2}{2} K\left(\frac{1+y}{2}\right)\right]\,,
\ee
and relating $y=\frac{1}{\sqrt{1+\frac{\Delta^2}{\mu^2}}}$\,. It turns out that the condition $g'(y)=0$ is equivalent to
\be
K\left(\frac{1+y}{2}\right)=2 E\left(\frac{1+y}{2}\right)\,,
\ee
which is not easy to solve in closed form, but gives $y\simeq 0.65223\ldots$ or
\be
\lim_{a_0\rightarrow -\infty}\Delta=\mu\times 1.1622\ldots\,,
\ee
for the gap parameter.

By contrast, in the weak coupling case, the condition $\frac{d p}{d \Delta}=0$ forces $\Delta$ to be very small. Therefore, expanding the elliptic integrals for small $\Delta$, we have
\be
\Delta^2=64 \mu^2\times e^{-4+\frac{2 \sqrt{2}\pi^2}{C_0 M^\frac{3}{2}\sqrt{\mu}}}\,,
\ee
which upon using (\ref{c0def}) gives the desired expression for the gap $\Delta$ in the weak coupling limit. 

\subsection*{Solution Problem 2.2}

From (\ref{EoSLO}) we find that in the limit $a_0\rightarrow -\infty$ we have
\be
\epsilon=\frac{3}{5} \mu n\,.
\ee
Now using $n$ from (\ref{nnnLO}) to express $\mu$ in terms of $n$, this gives
\be
\lim_{a_0\rightarrow -\infty}\epsilon=\frac{3}{10 M} n^{\frac{5}{3}} \left(\frac{3\pi^2}{g(y)}\right)^{\frac{2}{3}}\,.
\ee
Using finally the value $y\simeq 0.65223$ from problem 2.1, and evaluating the function $g(y)$ for this argument gives $g(y)\simeq 2.2032$, so that
\be
\xi\simeq 0.590606\ldots
\ee
for the Bertsch parameter.

\subsection*{Solution Problem 3.1}

We take the hint in the problem assignment and add a vacuum energy and (bare) mass term to the O(N) model Lagrangian, that is, instead of (\ref{4don}) we study the Euclidean action
\be
\label{4donM}
S_E=S_0+\int d^4x \left[\frac{1}{2}\partial_\mu \vec{\phi}\cdot \partial_\mu \vec{\phi}+\frac{m_0^2}{2}\vec{\phi}^2+\frac{\lambda_0}{N}\left(\vec{\phi}^2\right)^2\right]\,,
\ee
where $S_0,m_0,\lambda_0$ are the bare parameters of the theory. The introduction of the auxiliary field progresses along the same lines as in lecture 1,2 and 3, and after dropping the large N suppressed auxiliary field fluctuations, we have
\be
\lim_{N\gg 1}Z=\int d\zeta_0 e^{-S_0-\frac{N}{2} {\rm Tr}\ln\left[-\partial_\mu \partial_\mu+2 i \zeta_0+m_0^2\right]-\frac{N}{4\lambda_0}\int d^4x \zeta_0^2}\,.
\ee

The integral we need to regularize is
\be
\frac{1}{2{\rm vol}}{\rm Tr}\ln\left[-\partial_\mu \partial_\mu+m^2\right]=J(m)=\frac{1}{2}\int \frac{d^4k}{(2\pi)^4}\ln\left(k^2+m^2\right)\,,
\ee
where ${\rm vol}=\int d^4x$ and $m^2=2i\zeta_0+m_0^2$ is now a combination of auxiliary zero mode and bare mass parameter.  Since the integrand has spherical symmetry, we can split
\be
J(m)=\frac{1}{2 (2\pi)^4}\int d\Omega \int_0^{\Lambda_{\rm UV}} dk\, k^3 \ln\left(k^2+m^2\right)\,,
\ee
where we have put a cut-off $\Lambda_{\rm UV}$ on the momentum integration to regularize the divergence. The integral over the solid angle $\Omega$ in $d$ dimensions is known as
\be
\int d\Omega=\frac{2 \pi^{\frac{d}{2}}}{\Gamma\left(\frac{d}{2}\right)}\,,
\ee
so that for $d=4$ we have
\be
J(m)=\frac{1}{(4 \pi)^2} \int_0^{\Lambda_{\rm UV}} dk\, k^3 \ln\left(k^2+m^2\right)\,.
\ee
The integral over momenta can be performed in closed form, and for $\Lambda_{\rm UV} \gg m^2$ we get
\be
J(m)=\frac{1}{64\pi^2}\left[\Lambda_{\rm UV}^4\ln \left(\Lambda_{\rm UV}^2 e^{-\frac{1}{2}}\right)+\Lambda_{\rm UV}^2 m^2+m^4 \ln\left(\frac{m^2}{\Lambda_{\rm UV}^2}\right)\right]\,.
\ee
Therefore,
\be
\lim_{N\gg 1}Z=\int d\zeta_0 e^{-S_0-\frac{N {\rm vol}}{64\pi^2}\left[\Lambda_{\rm UV}^4\ln \left(\Lambda_{\rm UV}^2 e^{-\frac{1}{2}}\right)+\Lambda_{\rm UV}^2 m^2+m^4 \ln\left(\frac{m^2}{\Lambda_{\rm UV}^2}\right)+\frac{16 \pi^2 \zeta_0^2}{\lambda_0}\right] }\,,
\ee
where $m^2=2 i \zeta_0+m_0^2$.

The exponent contains terms in $m$ and $\zeta_0$, which are not independent. So we rewrite
\be
\zeta_0^2=-\frac{1}{4}(2 i \zeta_0)^2=-\frac{1}{4}(m^2-m_0^2)^2=-\frac{1}{4}(m^2-m_0^2)^2=-\frac{m^4}{4}-\frac{m_0^4}{4}+\frac{m^2 m_0^2}{2}\,,
\ee
such that the exponent contains the bare parameters $m_0,\lambda_0,S_0$ and the integration variable $m$.

Now let's renormalize: we choose the vacuum energy constant $S_0$ such that it cancels the terms in the exponent that do not depend on $m$ (obviously, since $S_0$ is a constant and $m$ is still being integrated over, so we cannot absorb any $m$-dependent terms in a constant):
\be
S_0=-\frac{N {\rm vol}}{64\pi^2} \left[\Lambda_{\rm UV}^4\ln \left(\Lambda_{\rm UV}^2 e^{-\frac{1}{2}}\right)-\frac{4 \pi^2 m_0^4}{\lambda_0}\right]\,.
\ee
This leaves us with
\be
\lim_{N\gg 1}Z=\int d\zeta_0 e^{-\frac{N {\rm vol}}{64\pi^2}\left[\Lambda_{\rm UV}^2 m^2+m^4 \ln\left(\frac{m^2}{\Lambda_{\rm UV}^2}\right)-\frac{4 \pi^2 m^4}{\lambda_0}+\frac{8 \pi^2 m^2 m_0^2}{\lambda_0}\right] }\,.
\ee

Next, renormalize the bare coupling by ensuring that the term proportional to $m^4$ in the exponent is finite. This can be done by absorbing the logarithmic UV-divergence into $\lambda_0$:
\be
\label{choice}
\ln\left(\frac{\mu^2}{\Lambda_{\rm UV}^2}\right)-\frac{4\pi^2}{\lambda_0}=-\frac{4\pi^2}{\lambda_{R}(\mu)}\,,
\ee
where we had to introduce a fictitious renormalization scale $\mu$ so that the argument of the logarithm does not have dimensions. We then have
\be
\lim_{N\gg 1}Z=\int d\zeta_0 e^{-\frac{N {\rm vol}}{64\pi^2}\left[\Lambda_{\rm UV}^2 m^2+m^4 \ln\left(\frac{m^2}{\mu^2}\right)-\frac{4 \pi^2 m^4}{\lambda_R(\mu)}+\frac{8 \pi^2 m^2 m_0^2}{\lambda_0}\right] }\,.
\ee

From (\ref{choice}), we see that
\be
\lambda_R(\mu)=\frac{1}{\frac{1}{\lambda_0}+\frac{1}{4\pi^2}\ln\frac{\Lambda_{\rm UV}^2}{\mu^2}}\,,
\ee
and we have a choice for $\lambda_0$. Choosing $\lambda_0>0$ leads to a quantum trivial theory. However, choosing $\lambda_0<0$, e.g.
\be
\lambda_0=\frac{4\pi^2}{\ln\frac{\Lambda^2_{\overline {\rm MS}}}{\Lambda_{\rm UV}^2}}\,,\quad \Lambda_{\overline {\rm MS}}<\Lambda_{\rm UV}\,,
\ee
leads to the running coupling result (\ref{runningL}) in lecture 3.

Finally, cancel the remaining UV-divergence in the term proportional to $m^2$: choosing 
\be
\Lambda_{\rm UV}^2+\frac{8 \pi m_0^2}{\lambda_0}=\frac{8 \pi m_R^2(\mu)}{\lambda_R(\mu)}\,,
\ee
achieves this, and we are left with
\be
\lim_{N\gg 1}Z=\int d\zeta_0 e^{-\frac{N {\rm vol}}{64\pi^2}\left[m^4 \ln\left(\frac{m^2}{\mu^2}\right)-\frac{4 \pi^2 m^4}{\lambda_R(\mu)}+\frac{8 \pi^2 m^2 m_R^2(\mu)}{\lambda_R(\mu)}\right] }\,.
\ee

All terms in the exponent are now finite, and we can proceed to perform the integral over $\zeta_0$. In particular, choosing $m_R(\mu)=0$ (which corresponds to $m_0^2=-\frac{\lambda_0 \Lambda_{\rm UV}^2}{8\pi}>0$ for $\lambda_0<0$) brings us back to the result in dimensional regularization used in lecture 3.

\subsection*{Solution Problem 3.2}

The finite-temperature calculation proceeds exactly along the same steps as in lecture 3, resulting in the large N partition function \ref{L3z1}. The only difference is that now the functional trace has to take into account the fact that our volume is the thermal cylinder, with the Euclidean time in the compact interval $\tau\in [0,\beta]$ and periodic boundary conditions. We have
\be
Z=\int d\zeta_0 e^{-N {\rm vol}\times V_{\rm eff}(\sqrt{2 i \zeta_0})}\,,
\ee
where
\be
V_{\rm eff}(m)=J(m)-\frac{m^4}{16\lambda}\,,
\ee
where $J(m)$ now is given in terms of a sum-integral:
\be
J(m)=\frac{1}{2{\rm vol}}{\rm Tr}\ln\left[-\partial_\mu\partial_\mu+m^2\right]
=\frac{T}{2}\sum_{n}\int \frac{d^{3-2\varepsilon}}{(2\pi)^{3-2\varepsilon}} \ln[\omega_n^2+k^2+m^2]\,,
\ee
with $\omega_n=2 \pi n T$ the bosonic Matsubara frequencies.
The large N evaluation of $Z$  leads to the saddle point condition
\be
\frac{dV_{\rm eff}(m)}{dm^2}=0\,,
\ee
of which $\frac{dJ}{dm^2}$ is one part. Specifically, we have
\be
0=\frac{d V_{\rm eff}(m)}{dm^2}=\frac{dJ(m)}{dm^2}-\frac{d}{dm^2}\frac{m^4}{16\lambda}\,.
\ee
So we need
\be
\frac{d J(m)}{dm^2}=\frac{T}{2}\sum_n \int_{\bf k}\frac{1}{\omega_n^2+k^2+m^2}=\int \frac{d^{3-2\varepsilon}}{(2\pi)^{3-2\varepsilon}}\frac{{\rm coth}\frac{\sqrt{k^2+m^2}}{2 T}}{2 \sqrt{k^2+m^2}}\,.
\ee
Using $\coth\frac{x}{2 T}=1+2 n_B(x)$ with $n_B(x)=\frac{1}{e^{x/T}-1}$, we have
\be
\frac{d J(m)}{dm^2}=\int_{\bf k} \frac{1}{2 \sqrt{k^2+m^2}}+\int_{\bf k}\frac{n_B(\sqrt{k^2+m^2})}{\sqrt{k^2+m^2}}\,.
\ee
Since the second term in this expression vanishes in the zero temperature limit $T\rightarrow 0$, the first term must be the zero temperature contribution we already calculated in the lecture. Specifically, using (\ref{id0}), we have
\be
\frac{d J_{T=0}(m)}{dm^2}\int_{\bf k} \frac{1}{2 \sqrt{k^2+m^2}}=-\frac{m^2}{32 \pi^2}\left(\frac{1}{\varepsilon}+\ln\frac{\bar\mu^2 e^1}{m^2}\right)\,.
\ee
We can evaluate the thermal piece by expanding the Bose-Einstein distribution function:
\be
\int \frac{d^3k}{(2\pi)^3}\frac{n_B(\sqrt{k^2+m^2})}{\sqrt{k^2+m^2}}=\sum_{n=1}^\infty \int \frac{d^3k}{(2\pi)^3} \frac{e^{-n \beta \sqrt{k^2+m^2}}}{\sqrt{k^2+m^2}}
=\sum_{n=1}^\infty\frac{m T}{2\pi^2 n}K_1(n \beta m)\,.
\ee
Putting everything together, we have the finite-temperature saddle-point condition
\be
0=-\frac{m^2}{32\pi^2}\left(\frac{1}{\varepsilon}+\ln\frac{\bar\mu^2 e^{1}}{m^2}\right)+\sum_{n=1}^\infty \frac{m T K_1(n \beta m)}{2 \pi^2 n}-\frac{m^2}{8 \lambda}\,.
\ee

The bare coupling constant is renormalized as before, so that we get for the finite-temperature saddle point condition
\be
\label{finiteTsaddle}
0=-\frac{m^2}{32\pi^2}\ln \frac{\Lambda^2_{\overline{\rm MS}}e^{1}}{m^2}+m T \sum_{n=1}^\infty \frac{K_1(n \beta m)}{2 \pi^2 n}\,.
\ee
At zero temperature, we recognize the perturbative saddle $m=0$, and the non-perturbative saddle at $m=\sqrt{e}\Lambda_{\overline{\rm MS}}$. At high temperature, the sum over Bessel function dominates, and there is no real solution to (\ref{finiteTsaddle}). At $T_c>T\neq 0$, we have two real solutions fulfilling
\be
\label{fTsaddle}
m\ln \frac{\Lambda^2_{\overline{\rm MS}} e^1}{m^2}=8 T \sum_{n=1}^\infty \frac{K_1(n \beta m)}{n}\,.
\ee
For fixed $T$, the sum over Bessel functions is a monotonically decreasing function of $m$, whereas the logarithm is a function that first rises, has a maximum, and then decreases and becomes negative for large $m$. The critical temperature $T_c$ is the one where the sum over Bessel functions just touches the logarithm function at one value of $m$ only.

Solving (\ref{fTsaddle}) numerically, one finds that this is the case for
\be
T=T_c\simeq 0.615663\ldots\times \Lambda_{\overline{\rm MS}}\,.
\ee

\subsection*{Solution Problem 3.3}

We'll need the perturbative running coupling, which in $\overline{\rm MS}$ scheme is given by
\be
\label{runningas}
\frac{d a_s}{d\ln \bar\mu^2}=-\beta_0 a_s^2-\beta_1 a_s^3-\beta_2 a_s^4-\ldots\,,
\ee
where \cite{vanRitbergen:1997va}
\be
a_s\equiv \frac{\alpha_s}{4\pi}\,,\quad \beta_0=11-\frac{2}{3}N_f\,,\quad
\beta_1=102-\frac{38}{3}N_f\,,\quad \beta_2=\frac{2857}{2}-\frac{5033}{18}N_f+\frac{325}{54}N_f^2\,.
\ee
Using the particle data book world average value
\be
\alpha_s(M_Z)=0.1179\pm 0.0009\,,
\ee
where $M_Z=91.1876$ GeV is the Z-mass, one can solve the running (\ref{runningas}) numerically down from $\bar\mu=M_Z$ until $\alpha_s$ diverges. For the various loop orders, this then leads to the following values of $\Lambda_{\overline{\rm MS}}$ for $N_f=5$:

\begin{centering}
  \hspace*{5cm}
\begin{tabular}{cc}
  Loop order & $\Lambda_{\overline{\rm MS}}$\\
  1 & 87 MeV\\
  2 & 242 MeV\\
  3 & 288 MeV
\end{tabular}
\end{centering}

While the result has not converged yet, we can try to use the 3-loop result
\be
\Lambda_{\overline{\rm MS}}\simeq 288 MeV\,,
\ee
as an estimate for the Landau pole in QCD. According to the formula (\ref{tc}), this then implies a critical temperature of
\be
T_c\simeq 177 {\rm MeV}\,,
\ee
for QCD. This is surprisingly close to
\be
T_c=176\pm 7 {\rm MeV}\,,
\ee
found for the confinement-deconfinement transition from lattice QCD \cite{Aoki:2006br}.

\bibliography{PT}
\end{document}